\catcode`\@=11
\def\citen#1{\if@filesw \immediate\write \@auxout {\string\citation{#1}}\fi%
\@tempcntb\m@ne \let\@h@ld\relax \def\@citea{}%
\@for \@citeb:=#1\do {\@ifundefined {b@\@citeb}%
    {\@h@ld\@citea\@tempcntb\m@ne{\bf ?}%
    \@warning {Citation `\@citeb ' on page \thepage \space undefined}}%
    {\@tempcnta\@tempcntb \advance\@tempcnta\@ne
    \setbox\z@\hbox\bgroup\ifcat0\csname b@\@citeb \endcsname \relax
    \egroup \@tempcntb\number\csname b@\@citeb \endcsname \relax
    \else \egroup \@tempcntb\m@ne \fi \ifnum\@tempcnta=\@tempcntb
    \ifx\@h@ld\relax \edef \@h@ld{\@citea\csname b@\@citeb\endcsname}%
    \else \edef\@h@ld{\hbox{--}\penalty\@highpenalty
    \csname b@\@citeb\endcsname}\fi
    \else \@h@ld\@citea\csname b@\@citeb \endcsname \let\@h@ld\relax \fi}%
\def\@citea{,\penalty\@highpenalty\hskip.13em plus.13em minus.13em}}\@h@ld}
\def\@citex[#1]#2{\@cite{\citen{#2}}{#1}}%
\def\@cite#1#2{\leavevmode\unskip\ifnum\lastpenalty=\z@\penalty\@highpenalty\fi%
  \ [{\multiply\@highpenalty 3 #1%
  \if@tempswa,\penalty\@highpenalty\ #2\fi}]}   %
\makeatother 

\newcommand\ad[1]  {{\rm ad}_{#1}^{}}
\def\adxt          {\ad X}
\newcommand\adxT[1]{{\rm ad}_\xt^{#1}}

\def\alg           {algebra}
\def\alphab        {{\Bar\alpha}}
\def\alphav        {{\Bar\alpha}^\Vee}
\def\alphaV        {({\Bar\alpha}^\Vee)}
\def\autg          {{\rm Aut}(\g)}
\def\auto          {automorphism}
\def\Bar           {\bar}  
\def\be            {\begin{equation}}
\def\bearl         {\begin{array}{l}}
\def\bearll        {\begin{array}{ll}}
\def\bearlll       {\begin{array}{lll}}
\def\betab         {{\Bar\beta}}
\def\betj          {\vec{\Bar\gamma}_{(j)}}
\def\betk          {\vec{\Bar\gamma}_{(k)}}
\def\bfe           {{\bf1}}
\def\bl            {\beta}
\def\bla           {block algebra}
\def\Bntilde       {{\tilde B_n\twtw}}
\def\bp            {\mbox{${\liefont b}$}}

\def\calm          {{\cal M}}

\def\cb            {chiral block}
\def\cbb           {\mbox{$\cal B$}}      

\def\cbbl          {\mbox{${\cal B}_{\Vec\Lambda}$}}
\def\cbbol         {\mbox{${\cal B}_{\omdv\Vec\Lambda}$}}
\def\cbd           {\mbox{\sf B$^\dual$}} 
\def\cbs           {\mbox{\sf B}}         
\def\cbsl          {\mbox{\sf B$_{\Vec\Lambda}$}}
\def\cbsol         {\mbox{\sf B$_{\omdv\Vec\Lambda}$}}
\def\Cdot          {\,{\cdot}\,}
\def\CEG           {\mbox{$\complex_\epsilon(\GG/\GGO)$}}
\def\CEGC          {\mbox{${\cal Z}(\complex_\epsilon(\GG/\GGO))$}}
\def\CEL           {\mbox{$\complex_\epsilon(\GFL/\GGO)$}}
\def\CELC          {\mbox{${\cal Z}(\complex_\epsilon(\GFL/\GGO))$}}
\def\cft           {conformal field theory}
\def\Cft           {Conformal field theory}
\def\cfts          {conformal field theories}

\def\Circ          {\,{\circ}\,}
\newcommand\coi[2] {\lfloor #1\rfloor^{}_{#2}}

\def\coin          {co-in\-va\-ri\-ant}
\def\complex       {{\dl C}}
\def\Complex       {$\dl C$}
\def\Con           {Conformal }

\def\csa           {Cartan subalgebra}

\def\demuv         {\delta^{}_{\vec{\Bar\mu}}}
\def\Demuv         {\vec\delta^{}_{\vec{\Bar\mu}}}
\def\demuvo        {\delta^{}_{\vec{\Bar\mu};\so}}
\def\df            {\,{:=}\,}

\def\dfsO          {{\hat g_\so}}

\def\diff          {{\rm d}}
\def\dim           {{\rm dim}\,}
\def\dl            {\mathbb }

\def\dsp           {\partial_{s'}}
\def\dsum          {\displaystyle\sum}
\def\dsty          {\displaystyle}
\def\dual          {\star}
\newcommand\e[2]   {e_{#1,#2}}
\newcommand\E[3]   {{\rm Ad}_{#3;#1,#2}}
\def\eal           {E^{\alphab}}
\def\ealbe         {E^{\alphab+\betab}}
\def\eals          {E^{\alphab_s}}
\def\ebe           {E^{\betab}}
\def\ebepm         {E^{\betab\pm\alphab}}
\def\ee            {\end{equation}}

\def\eear          {\end{array}}
\def\eigenspace    {invariant subspace}
\def\emal          {E^{-\alphab}}
\def\emals         {E^{-\alphab_s}}

\def\empal         {E^{\mp\alphab}}

\def\epmal         {E^{\pm\alphab}}
\def\eq            {\,{=}\,}
\newcommand\erf[1] {(\ref{#1})}
\newcommand\Erf[2] {(\ref{#1#2})}

\newcommand\expad[1]{\exp({\rm ad}_{#1}^{})}
\def\exs           {\exp(\half\ipi\,\Xs)}
\def\Exs           {\exp\llb\Frac{\ipi}2\,\Xs\lrb}
\def\exis          {\exp(\half\ipi\,X_{\alpha_{i_s}^{};s,\so})}
\def\exso          {\exp(\half\ipi\,\Xso)}
\def\Exsys         {\E{x_s}{y_s}{\ipi/2}}
\def\Exy           {\E{\Xs}{\Yss}{\ipi/2}}
\def\eyis          {\exp(-\half\ipi\,Y_{\alpha_{i_s}^{};\so})}
\def\eys           {\exp(-\half\ipi\,\Yss)}
\def\ezalbe        {E^{2\alphab+\betab}}

\def\fe            {f_{\sss+}}
\def\fE            {f_1}

\def\feS           {\varphi^{(s)}}

\def\feSp          {\varphi^{(s)}}
\def\feSP          {\varphi^{(s')}}
\def\fez           {{\fe\fz}}

\def\findim        {finite-dimensional}
\newcommand\fis[1] {\varphi^{(#1)}}
\def\fism          {\varphi_{m,s}}

\def\fiss          {\varphi_{\so,s}}
\def\fisso         {\varphi_{\so,s}^\kln}
\def\fissp         {\varphi_{\so,s'}}
\def\fisp          {\varphi_{s,s'}}
\def\fisP          {\varphi_{\so,s'}}
\def\fisu          {\varphi_{\so,u}}
\def\fisuo         {\varphi_{\so,u}^\kln}
\def\fiuP          {\varphi_{u,s'}}
\def\fius          {\varphi_{u,s}}
\def\fiuso         {\varphi_{u,s}^\kln}
\def\fmp           {f_{\sss\mp}}
\def\fpm           {f_{\sss\pm}}
\newcommand\Frac[2]{\mbox{\large$\frac{#1}{#2}$}}
\newcommand\fraC[2]{{#1}/{#2}}
\def\fs            {f\ss}
\def\fso           {f\sso}
\def\fsO           {{\tilde h_\so}}
\def\FsO           {{h_\so}}
\def\fss           {\varphi\sso^{(s)}}
\def\fsu           {{\tilde h_u}}
\def\fun           {\mbox{${\cal F}({\dl P}^1_\mm)$}}
\def\funz          {\mbox{${\cal F}^*({\dl P}^1_\mm)$}}
\def\futnote#1     {\footnote{~#1}\ }
\def\fz            {f_{\sss-}}
\def\fZ            {f_2}
\def\g             {\mbox{$\liefont g$}}
\def\G             {\mbox{$\Gamma_{\!\!{\rm w}}$}}
\def\gammav        {{\vec{\Bar\gamma}}}
\def\gb            {\mbox{$\bar\gM$}}
\def\gB            {{\bar\gM}}
\def\gE            {\,{\ge}\,}
\def\GF            {\mbox{$\Gamma_{\!\!{\rm fix}}$}}
\def\GFl           {{\Gamma_{\!\vec\Lambda}}}
\def\GFL           {\mbox{$\Gamma_{\!\!\vec\Lambda}$}}
\def\GG            {\Gamma_{\!\!{\rm w}}}
\def\GGO           {\Gamma}
\def\gloopb        {\mbox{$\bar\gM_{\rm loop}$}}

\def\gM            {{\liefont g}}

\def\gmp           {g_\mp}
\def\go            {\mbox{$\gM^{}_\circ$}}
\def\GO            {\mbox{$\Gamma$}}

\def\gP            {g_{\sss+}}
\def\gpm           {g_\pm}
\def\GQ            {\mbox{$\Gamma_{\!\!{\rm out}}$}}
\def\gs            {{\gM_{s}}}
\def\gS            {\mbox{$\gM_{s}$}}
\def\gsauto        {mul\-ti-shift automorphism}
\def\Gsauto        {Mul\-ti-shift automorphism}
\def\gso           {\mbox{$\gM_\so$}}
\def\gsO           {{g_\so}}
\def\gsosp         {{g_{\so,s'}}}
\def\gsu           {{g_{\so,u}}}
\def\gsos          {{g_{\so,s}}}
\def\gsou          {{g_{\so,u}}}
\def\gt            {{\bar\gM{\otimes}{\cal F}}}
\def\gT            {\mbox{$\bar\gM{\otimes}{\cal F}$}}
\def\gus           {{g_{u,s}}}
\def\gV            {{g^\Vee}}

\def\js            {\mbox{$\omega_{\rm s}$}}
\def\jv            {\mbox{$\omega_{\rm v}$}}

\def\half          {\mbox{$\frac12\,$}}

\def\Halfkilling   {(\alphab,\alphab)^{-1}\,}
\def\hals          {H^{\alphab_s}}
\def\hbe           {H^{\betab}}
\newcommand\hepv[1]{}
\def\hi            {H^i}
\def\hil           {\mbox{$\cal H$}}

\def\hl            {\mbox{${\cal H}_\Lambda$}}

\def\hs            {{\tilde g_s}}
\def\hsa           {horizontal subalgebra}
\newcommand\hsp[1] {\mbox{\hspace{#1 em}}}
\def\hw            {highest weight}

\def\hwv           {highest weight vector}
\def\hy            {$\mbox{-\hspace{-.66 mm}-}$}
\def\id            {{\rm id}}
\def\iD            {{1}}
\def\ii            {{\rm i}}
\def\ihwm          {irreducible highest weight module}
\def\infdim        {infinite-dimensional}
\def\iN            {\,{\in}\,}
\newcommand\Inner[1]{A_{#1;\so}}
\newcommand\INner[1]{A_{#1}}
\def\intg          {{\rm Int}(\g)}
\def\ipi           {{\rm i}\pi}

\def\killing       {\Frac2{(\alphab,\alphab)}}
\def\killings      {\Frac2{(\alphab_s,\alphab_s)}}
\def\kk            {\mbox{$k$}}
\def\kln           {{\scriptscriptstyle[0]}}

\def\Kln           {{\scriptscriptstyle\,[0]}}

\def\kma           {Kac\hy Moo\-dy algebra}
\def\kpf           {Kac\hy Peterson formula}
\def\kv            {\mbox{$k^\Vee$}}
\def\kV            {{k^\Vee}}

\def\kzc           {Knizh\-nik\hy Za\-mo\-lod\-chi\-kov connection}

\newcommand\La[1]  {{\Lambda_{#1}}}
\long\def\labl#1   {\label{#1}\ee}
\long\def\Labl#1#2 {\label{#1#2}\ee}
\def\lambdab       {{\Bar\lambda}}
\def\Lambdab       {{\Bar\Lambda}}
\def\Lao           {{\Lambda_\circ}}
\def\Laod          {{\Lambda_\circ^+}}
\def\Lap           {{\Lambda'}}
\def\Lapp          {{\Lambda''}}
\def\Las           {{\La s}}
\def\Ldots         {,...\,,}
\def\leu           {L_{-1}^{\scriptscriptstyle(u)}}
\def\lhs           {left hand side}

\def\lie           {Lie algebra}

\def\liefont       {\mathfrak }

\def\lifo          {linear form}
\def\lLb           {\mbox{\large[}}
\def\llb           {\mbox{\large(}}
\def\LLb           {\mbox{\Large[}}
\def\Llb           {\mbox{\Large(}}
\def\LlB           {\mbox{\LARGE(}}
\def\lr            {\mbox{$L$}}
\def\lRb           {\mbox{\large]}}
\def\lrb           {\mbox{\large)}}
\def\LRb           {\mbox{\Large]}}
\def\Lrb           {\mbox{\Large)}}
\def\LrB           {\mbox{\LARGE)}}
\def\lT            {\,{<}\,}
\def\lv            {\mbox{$L^{\!\VEE}_{\phantom|}$}}
\def\lw            {\mbox{$L^{}_{\rm w}$}}
\def\lwv           {\mbox{$L^{\!\VEE}_{\rm w}$}}

\def\me            {{m{-}1}}
\def\mi            {\,{-}\,}
\def\Mid           {\,{\mid}\,}
\def\mm            {{\!m}}
\def\mod           {\;{\rm mod}\,}
\def\MU            {{\rm M}}
\def\mub           {{\Bar\mu}}
\def\mus           {{\Bar\mu_s}}
\def\muso          {{\Bar\mu_{s_\circ}}}
\def\musp          {{\Bar\mu_s}}
\def\musP          {{\Bar\mu_{s'}}}
\def\muu           {{\Bar\mu_u}}
\def\muv           {{\vec{\Bar\mu}}}
\def\muvi          {\vec{\Bar\kappa}_{(i)}}
\def\muvim         {\vec{\Bar\kappa}_{(i(\muv))}}
\def\muvin         {\vec{\Bar\kappa}_{(i(\nuv))}}
\def\muvj          {\vec{\Bar\kappa}_{(j)}}
\def\nabls         {\nabla_{\!\!s}}
\def\nablu         {\nabla_{\!\!u}}

\def\nE            {\,{\not=}\,}
\def\none          { \{0\} }

\def\nub           {{\Bar\nu}}
\def\nus           {{\Bar\nu_s}}
\def\nuv           {{\vec{\Bar\nu}}}
\newcommand\nxl[1] {\\{}\\[-.#1em]}
\newcommand\Nxl[2] {\\{}\\[-#1.#2em]}
\def\omd           {\omega^\dual}
\def\omdv          {\vec\omega^\dual}
\def\Omeuv         {\vec\omega^{}_{\vec{\Bar\mu}}}
\def\omeuvs        {\omega^{}_{\vec{\Bar\mu};s}}
\def\omeuvsd       {\omega^\dual_{\vec{\Bar\mu};s}}
\def\ommuc         {[\ommuv]}
\def\ommucs        {[\sigma^{}_{\vec{\Bar\mu};s}]}

\def\ommuo         {\sigma^{\sss(0)}_{\Bar\mu}}
\def\ommuos        {\sigma^{\sss(0)}_{\Bar\mu_s}}

\def\omeus         {\omega^{}_{\Bar\mu_s}}
\def\ommuv         {\sigma^{}_{\vec{\Bar\mu}}}
\def\ommuV         {\sigma_{\vec{\Bar\mu}}}
\def\Ommuv         {\vec\sigma^{}_{\vec{\Bar\mu}}}
\def\ommuvm        {\sigma^{}_{\vec{\Bar\mu};m}}
\def\ommuvo        {\sigma^{}_{\vec{\Bar\mu};\so}}
\def\ommuvs        {\sigma^{}_{\vec{\Bar\mu};s}}
\def\ommuvu        {\sigma^{}_{\vec{\Bar\mu};u}}
\def\omo           {\omega_n^{}}
\def\omO           {\omega_n}
\def\Oms           {\omega^{}_{s}}

\def\onetom        {1,2,...\,,m}
\def\onetome       {1,2,...\,,\me}
\def\onetor        {1,2,...\,,r}
\newcommand\oo[3]  {{{#1}^{\mskip-#3 mu\raise #2 pt\hbox{$\scriptstyle\circ$}}}}
\def\oog           {\oo{\gM}{1.6}{10.3}}

\def\ot            {\raisebox{.07em}{$\scriptstyle\otimes$}}
\def\otim          {\ot}
\def\oT            {\,\ot\,}

\def\otimeS        {\,{\otimes}\,}

\def\outg          {{\rm Out}(\g)}
\newcommand\pair[2]{\langle #1\,{,}\,#2 \rangle}

\def\pe            {\mbox{${\dl P}^1$}}

\def\pl            {\,{+}\,}
\newcommand\prd[1] {\Phi_{#1}}
\def\produ         {\prod_{s=1\atop s\ne u}^m}

\def\REG           {(\GG/\GGO)^\circ_\epsilon}
\def\REGL          {(\GFl/\GGO)^\circ_\epsilon}
\def\rep           {rep\-re\-sen\-ta\-ti\-on}
\def\Rep           {Representation}
\newcommand\res[1] {\Res_{#1}}
\def\Res           {{\cal R}\!\mbox{\sl es}}

\def\resp          {respectively}

\def\rhs           {right hand side}
\def\rmd           {{\rm d}}
\def\rmdd          {{\rm d}}
\def\sauto         {shift automorphism}

\newcommand\sect[1] {\section{#1}\setcounter{equation}{0}}

\def\sicu          {\mbox{$\omega$}}
\def\sigmat        {\tilde\sigma}

\def\so            {{s_\circ}}

\def\splus         {\oplus} 
\def\ss            {_{|s}}
\def\sse           {^{\scriptscriptstyle(1)}}
\def\ssi           {^{\scriptscriptstyle(i)}}
\newcommand\ssn[1] {_{|#1}}
\def\sso           {_{|\so}}
\def\ssp           {_{|s'}}
\def\sss           {\scriptscriptstyle}

\def\sssp          {{\scriptscriptstyle(s')}}
\def\ssss          {{\scriptscriptstyle(s)}}
\def\ssz           {^{\scriptscriptstyle(2)}}
\def\sumlL         {\sum_{\ell,\ell'\in\zet}}
\def\sumLL         {\sum_{\!\!\ell,\ell'\in\zet}}
\def\sumlz         {\sum_{\ell\in\zet}}
\def\sumLz         {\sum_{\ell'\in\zet}}
\def\summs         {\sum_{s=1}^m}
\def\summso        {\sum_{\so=1}^m}
\def\summsp        {\dsum_{s=1}^m}
\def\summse        {\sum_{s=1}^\me}
\def\summu         {\dsum_{s=1\atop s\ne u}^m}
\newcommand\sumre[1]{{\displaystyle\sum_{#1=1}^r}}
\def\sumsge        {\dsum_{s=1\atop s>\so}^\me}
\def\sumss         {\dsum_{s=1\atop s\ne\so}^m}
\def\sumsse        {\dsum_{s=1\atop s\ne\so}^\me}
\def\Tau           {\Theta}
\def\Tausig        {\Theta_\sigma}

\def\Tilde         {{}}  
\def\Times         {\,{\times}\,}
\def\To            {\mapsto}

\def\Tommug        {\vec\Tau_{\vec{\Bar\gamma}}}
\def\TommugA       {\vec\Tau_{\vec{\Bar\gamma}_1}}
\def\TommugAB      {\vec\Tau_{\vec{\Bar\gamma}_1+\vec{\Bar\gamma}_2}}
\def\TommugB       {\vec\Tau_{\vec{\Bar\gamma}_2}}
\def\Tommugm       {\vec\Tau_{\muvim}}
\def\Tommugmn      {\vec\Tau_{\muvim+\muvin}}
\def\Tommugn       {\vec\Tau_{\muvin}}
\def\Tommugv       {\vec\Tau_{\vec{\Bar\gamma}(\muv)}}
\def\Tomnugv       {\vec\Tau_{\vec{\Bar\gamma}(\nuv)}}
\def\Tommnugv      {\vec\Tau_{\vec{\Bar\gamma}(\muv)+\vec{\Bar\gamma}(\nuv)}}
\def\tommuv        {\Tau_{\vec{\Bar\mu}}}
\def\Tommuv        {\vec\Tau_{\vec{\Bar\mu}}}
\def\TommuvA       {\vec\Tau_{\vec{\Bar\mu}_1}}
\def\TommuvAB      {\vec\Tau_{\vec{\Bar\mu}_1+\vec{\Bar\mu}_2}}
\def\TommuvB       {\vec\Tau_{\vec{\Bar\mu}_2}}

\def\Tommuvd       {\vec\Tau^\dual_{\vec{\Bar\mu}}}
\def\Tommuvdc      {\vec\Tau^\dual_{[\vec{\Bar\mu}]}}

\newcommand\tommuvI[1]{\Tau_{\vec{\Bar\mu};#1}}
\def\Tommuvi       {\vec\Tau_{\muvi}}

\def\Tommuvij      {\vec\Tau_{\muvi+\muvj}}

\def\Tommuvj       {\vec\Tau_{\muvj}}

\def\TommuvM       {\vec\Tau^{-1}_{\vec{\Bar\mu}}}
\def\tommuvo       {\Tau_{\vec{\Bar\mu};\so}}
\def\tommuvs       {\Tau_{\vec{\Bar\mu};s}}
\def\tommuvu       {\Tau_{\vec{\Bar\mu};u}}

\def\Tommuvz       {\vec\Tau_{\vec{\Bar\mu}}(\vz)}

\def\Tomnuv        {\vec\Tau_{\vec{\Bar\mu}+\vec{\Bar\nu}}}
\def\tomnuvo       {\Tau_{\vec{\Bar\mu}+\vec{\Bar\nu};\so}}

\def\Tonnuv        {\vec\Tau_{\vec{\Bar\nu}}}
\def\tonnuvo       {\Tau_{\vec{\Bar\nu};\so}}
\def\Tr            {{\rm T}}

\def\ttoalpho      {\tilde\Tau_{\alphav_s;\so}}
\def\ttobeto       {\tilde\Tau_{\betab^\Vee_{s'};\so}}
\def\ttommuv       {\Tilde\Tau_{\vec{\Bar\mu}}}

\def\Ttommuvj      {\Tilde{\vec\Tau}_{\vec{\Bar\gamma}_{(j)}}}

\def\Ttommuvk      {\Tilde{\vec\Tau}_{\vec{\Bar\gamma}_{(k)}}}
\def\ttommuvo      {\Tilde\Tau_{\vec{\Bar\mu};\so}}
\def\tttommuvo     {\tilde\Tau_{\vec{\Bar\mu};\so}}

\newcommand\twobrac[1]{(\!(#1)\!)}
\def\twodim        {two-di\-men\-si\-o\-nal}
\def\twtw          {^{(2)}}
\def\U             {{\sf U}}

\def\untw          {^{(1)}}
\def\Up            {\U^+\!}
\def\uz            {U^{}_{\so,s}(\vz)}
\def\Uz            {U^{}_{\so;\muv}(\vz)}
\def\uza           {U^{}_{\so,s;\alphav_s}(\vz)}
\def\uzm           {U^{-1}_{\so,s}(\vz)}
\def\uzo           {U_{\so,s}(\vzo)}
\def\uzz           {U^2_{\so,s}(\vz)}
\def\Uzz           {U^2_{\so;\muv}(\vz)}

\def\vac           {\Omega}
\def\Vec           {\vec}
\def\Vee           {{\scriptscriptstyle\vee}}
\def\VEE           {{\scriptscriptstyle\vee}_{\phantom|}}

\def\vir           {\mbox{${\cal V}\!${\sl ir}}}
\def\vira          {Virasoro algebra}

\def\vmub          {\vec{\Bar\mu}}
\def\vh            {{\Vec{\hil}}}
\def\vhd           {{\Vec{\hil}}_{}^\dual}
\def\vhl           {\hil_{\vec\Lambda}}
\def\vhol          {\hil_{\omdv\vec\Lambda}}
\def\vp            {{\Vec p}}
\def\vpo           {{\Vec p}^\Kln}
\def\vz            {{\Vec z}}
\def\vzo           {{\Vec z}^\Kln}

\def\wolog         {without loss of generality}

\def\wrt           {with respect to }
\def\wrtt          {with respect to the }
\def\WZW           {Wess\hy Zumino\hy Witten}
\def\wzwm          {WZW model}
\def\wzwt          {WZW theory}
\def\wzwts         {WZW theories}
\def\Xas           {{X_{\so,s;\alpha_s}}}
\def\xb            {{\Bar x}}
\def\xaf           {{X_{\alphab,f}}}
\def\xafe          {{X_{\alphab,\fE}}}
\def\xafz          {{X_{\alphab,\fZ}}}

\def\xs            {{x_{s}}}
\def\Xs            {{X_{\so,s}}}
\def\Xso           {{X_{\so,s}^\kln}}
\def\xt            {X}
\def\yb            {{\Bar y}}
\def\ys            {{y_{s}}}
\def\Ys            {{Y_{-\vec{\Bar\mu};s}}}
\def\Yss           {{Y_{s}}}
\def\yt            {Y}
\def\Yu            {{Y_{\vec{\Bar\mu};u}}}
\newcommand\z[1]   {\zeta_{#1}}
\def\zet           {{\dl Z}}

\def\zo            {\!\raisebox{-.23em}{\Large$|$}\raisebox{-.53em}
                   {$\scriptstyle\vz=\vzo$}}
\def\zuo           {\!\raisebox{-.23em}{\Large$|$}\raisebox{-.53em}
                   {$\scriptstyle z_u=z_u^\kln$}}

\catcode`\@=12

\documentclass[12pt]{article} \usepackage{amssymb,amsfonts}

\begin{document}

\begin{flushright}  {~} \\[-15 mm]  {\sf hep-th/9805026} \\[1mm]
{\sf CERN-TH/98-145} 
\end{flushright}

\begin{center} \vskip 15mm
{\Large\bf THE ACTION OF OUTER AUTOMORPHISMS}\\[4mm]
{\Large\bf ON BUNDLES OF CHIRAL BLOCKS}\\[16mm]
{\large J\"urgen Fuchs} \\[3mm]
Max-Planck-Institut f\"ur Mathematik \\[.6mm]
Gottfried-Claren-Str.\ 26, \  D -- 53225~~Bonn \\[11mm]
{\large Christoph Schweigert} \\[3mm] CERN \\[.6mm] CH -- 1211~~Gen\`eve 23
\end{center}
\vskip 20mm
\begin{quote}{\bf Abstract}\\[1mm]
On the bundles of WZW chiral blocks over the moduli space of a punctured
rational curve we construct isomorphisms that implement the action of outer 
automorphisms of the underlying affine \lie. These bundle-isomorphisms 
respect the \kzc\ and have finite order. When all primary fields are fixed 
points, the isomorphisms are endomorphisms; in this case, the bundle of chiral 
blocks is typically a reducible vector bundle.
A conjecture for the trace of such endomorphisms is presented;
the proposed relation generalizes the Verlinde formula.
Our results have applications to \cfts\ based on non-simply connected groups
and to the classification of boundary conditions in such theories.
\end{quote}
\newpage


\sect{Introduction and summary}

\subsection{Chiral blocks}

Spaces of \cb s \cite{fefk} are of considerable 
interest both in physics and in mathematics. In this paper we construct 
a natural class of linear maps on the spaces of \cb s of WZW \cfts\
and investigate their properties. Recall that a \wzwt\ is a \twodim\ \cft\
which is characterized by an untwisted affine \lie\ \g\ and a positive
integer, the level.

The \cb\ spaces of a rational \cft\ are \findim\ complex vector spaces.
In the WZW case, one associates to each complex curve $C$ of genus $g$ with $m$
distinct smooth marked points and to each $m$-tuple $\vec\Lambda$ of integrable 
weights of \g\ at level $\kv$ a \findim\ vector space \cbsl\ of \cb s. 
These spaces are the building blocks for the correlation functions of
\wzwts\ both on surfaces with and without boundaries, and on orientable as 
well as unorientable surfaces (see e.g.\ \cite{fuSc6}).
Moreover, since \wzwts\ underly the construction of various other classes of 
\twodim\ \cfts, the spaces of WZW \cb s enter in the description of the 
correlation functions of those theories as well. They also form the
space of physical states in certain three-dimensional topological field
theories \cite{witt27}. Finally they are of interest to algebraic geometers 
because via the Borel\hy Weil\hy Bott theorem they are closely related to 
spaces of holomorphic sections in line bundles over
moduli spaces of flat connections (for a review, see e.g.\ \cite{beau2}).

It is necessary (and rather instructive) not to restrict ones attention to 
the case of fixed insertion points, but rather to vary both the moduli of the 
curve $C$ and the position of the marked points. In \cft\
this accounts e.g.\ for the dependence of correlation functions on the 
positions of the fields.)
This way one obtains a complex vector space \cbsl\
over each point in the moduli space $\calm_{g,m}$ of complex curves of genus 
$g$ with $m$ marked smooth points. It is known (see e.g.\ \cite{tsuy,Ueno})
that these vector spaces fit together into a vector bundle \cbbl\
over $\calm_{g,m}$, and that this vector bundle
is endowed with a projectively flat connection, the \kzc.
(Sometimes the term `\cb' is reserved for sections in the bundle \cbbl\ that
are flat with respect to the \kzc.) 
In this paper, we will restrict ourselves mainly to curves of genus $g\eq0$, 
but allow for arbitrary number $m\gE2$ of marked points. For brevity, we 
will denote the moduli space of $m$ distinct points on \pe\ simply by 
$\calm_m$. 

\subsection{Automorphisms}

It has been known for a long time that certain outer
automorphisms of the affine \lie\ \g\ that underlies a \wzwt\
play a crucial role in the construction
of several classes of \cfts. The automorphisms in question
are in one-to-one correspondence with the elements of the center 
$\cal Z$ of the relevant compact Lie group,
i.e.\ of the real compact connected and simply connected Lie group $G$ whose
\lie\ is the compact real form of the \hsa\ \gb\ of the affine \lie. 
These automorphisms underly the existence
of non-trivial modular invariants, both of extension and of automorphism type,
which describe the `non-diagonal' \wzwts\ that are based \cite{fegk}
on non-simply connected quotients of the covering group $G$.
(For the modular matrices of such theories see \cite{fusS6}.)
The same class of automorphisms also plays a crucial role in the construction
of `gauged' \wzwm s \cite{gaku3,kaSc2,hori}, i.e.\ coset \cfts\ \cite{fusS4}. 

Let us remark that the structures implied by such
automorphisms have been generalized for arbitrary rational \cfts\ in the theory
of so-called simple currents (for a review, see \cite{scya6}).

It is also worth mentioning
that in the representation theoretic approach to WZW \cb s \cite{Ueno,beau}
one heavily relies on the loop construction of the
affine \alg\ \g. That description treats the simple roots of the \hsa\ \gb\
and the additional simple root of \g\ on a rather different footing and
accordingly does not reflect the full structure of \g\ in a symmetric way. In 
particular the symmetries of the Dynkin diagram of \g\ which are associated to 
the (classes of) outer \auto s of \g\ that correspond to the center $\cal Z$
do not possess a particularly nice realization in the loop construction.

\subsection{Main results}

The basic purpose of this paper is to merge the two issues of \cb\ spaces and
outer \auto s of affine \lie s. Thus our goal is to implement and
describe the action of the outer automorphisms corresponding to the center
$\cal Z$ of $G$ on the spaces of \cb s of \wzwts. This constitutes in
particular a necessary input for the definition of
chiral blocks for the theories based on non-simply connected quotient groups 
of $G$. It also turns out to be an important
ingredient in the description of the possible boundary conditions of \wzwts\
with non-diagonal partition function \cite{fuSc5}. Moreover, our results
are also relevant for the quest of establishing a Verlinde formula (in the
sense of algebraic geometry) for moduli spaces of flat connections with
non-simply connected structure group.
(This might also pave the way to a rigorous proof of the Verlinde 
formula for more general \cft\ models, such as coset \cfts.)

Let us now summarize informally the main results of this paper.
Consider the direct sum of $m$ copies of the coweight lattice of \gb\
and denote by \G\ the subgroup consisting of those $m$-tuples $\muv\eq(\mub_i)$
which sum to zero, $\sum_{i=1}^m\mub_i=0$. To each $\muv\iN\G$
we associate a map $\omdv$ 
on the set of integrable weights at level $\kv$ and, for each 
$m$-tuple $\vec\Lambda$ of integrable weights, a map
  $$  \Tommuvd: \quad \cbbl \to \cbbol  $$  
which is an isomorphism of the bundles \cbbl\ of \cb s over the moduli space
$\calm_m$ and which projects to the identity on $\calm_m$. We show that this 
map respects the \kzc s on \cbbl\ and \cbbol. Finally, we show that the map 
$\Tommuvd$ is the identity map on \cbbl\ whenever the collection
$\muv$ of coweight lattice elements contains only elements of the coroot
lattice of \gb. We denote the sublattice of \G\ that
consists of such $m$-tuples $\muv$ by \GO. (With these properties the maps 
$\Tommuvd$ constitute prototypical examples
for so-called implementable automorphisms of the fusion rules. For the role
of such automorphisms in the determination of possible boundary conditions
for the \cft, see \cite{fuSc6,reSC}.)

Let us now discuss a few implications of the bundle-isomorphism $\Tommuvd$.
First, when $\omdv\vec\Lambda\nE\vec\Lambda$, then $\Tommuvd$ can be used to 
identify the two bundles \cbbl\ and \cbbol. 
On the other hand, suppose that the stabilizer \GFL, i.e.\ the subgroup of \G\
that consists of all elements which satisfy $\omdv\vec\Lambda\eq\vec\Lambda$,
is strictly larger than \GO. In this case the finite abelian group
$\GFL/\GO$ acts projectively by endomorphisms on each fiber. We then
decompose each fiber into the direct sum of eigenspaces under this action.
It is now crucial to note that the action of this finite group commutes with
the \kzc; this property implies that the eigenspaces fit together into
sub-vector bundles of \cbbl. In other words, the bundle of
\cb s is in this case a reducible vector bundle. Moreover, each
of these subbundles comes equipped with its own \kzc, which is
just the restriction of the \kzc\ of \cbbl\ to the subbundle.
This implies in particular that the chiral \cft\ that underlies the 
\wzwm\ with a modular invariant of extension type possesses a \kzc\ as well. 

This pattern -- identification, respectively splitting of fixed points into
eigenspaces -- is strongly reminiscent of what has been found in coset \cfts\ 
\cite{fusS4}. It is the deeper reason why similar formul\ae\ hold 
\cite{fusS6,fusS4} for the so-called resolution of fixed points in the 
definition of coset \cfts\ and in simple current modular invariants.

The rank of the subbundles is the same for all values of the moduli.
This rank is related via Fourier transformation over the subgroup $\REGL$ of 
regular elements of the finite abelian group
$\GFL/\GO$ to the traces of $\Tommuvd$ on any fiber. Thus we can conclude that 
these traces do not depend on the moduli either. We present a conjecture for 
a formula for these traces and support it by several consistency checks. 
We expect that several
features of our construction can be generalized to curves of arbitrary genus. 
So far we do not know how to prove our trace formul\ae. Some traces on spaces of
\cb s on higher genus surfaces with only the vacuum module involved have,
however, been computed \cite{beau3} in some special cases, and in these cases
they agree with our formula.

\subsection{Organization of the paper}

The rest of this paper is organized as follows. In section \ref{s.Au} 
we introduce (see equation \Erf Au)
a class of \auto s of an untwisted affine \lie\ \g, which we call
\gsauto s. These \auto s can be extended uniquely to the semidirect sum of \g\ 
with the Virasoro algebra; they depend 
on a collection of $m$ elements $\mus$ of the coweight lattice, and also on
a collection of pairwise distinct complex numbers $z_s$. We explain how
such automorphisms are implemented on irreducible highest weight modules
over the affine \lie\ and determine their class modulo inner automorphisms.

In section \ref{s.au} we introduce the algebra \gT\ of \lie\ valued meromorphic
functions over $\pe$ with possible finite order poles at $m$ prescribed 
different points, the so-called block algebra. The \lie\
\gT\ can be regarded as a subalgebra of $\g^m$, the
direct sum of $m$ copies of the affine \lie\ \g\ with all centers identified.
We observe that when the complex numbers $z_s$ that enter in the definition of 
the \gsauto s are interpreted as the coordinates of points on \pe, 
then a tensor product of the \auto s of section \ref{s.Au} restricts 
to an \auto\ of the block \alg, see formula \Erf aU. One should regard this 
latter \auto\ as a global object from which the former \auto s are obtained by 
local expansions.
In section \ref{s.II} we study the special case of \gsauto s that correspond
to elements in the coroot lattice. We show that they are inner automorphisms
with respect to the block algebra. 

In section \ref{s.cb} we combine the implementation of these automorphisms on 
the modules of the affine \lie\ at all insertion points,
so as to obtain an implementation on the modules over the \bla\ that arise from 
tensor products of the affine modules.
We show that for fixed values of the insertion points 
these maps induce isomorphisms on the spaces of \cb s and prove that 
(upon suitably fixing the phases of the implementing maps)
all these isomorphisms of \cb\ spaces have finite order. 

Up to this point the moduli of the problem, i.e.\ the positions of the 
punctures, are kept fixed in all our considerations. In section \ref{s.bu}
we proceed to consider the dependence of our constructions on these moduli. Our
guiding principle will be that the isomorphisms on the spaces of \cb s
are compatible with the \kzc. We show that the implementation on
the modules can indeed be chosen in a such way that the \kzc\ is preserved
and that for each value of the moduli the implementation on the module
provides the same projective \rep\ of the group \G.
We then show that for all values of the moduli the action of a \gsauto\ on the
\cb s is the identity map when all coweights are actually coroots, so that 
in this case the implementation $\Tommuvd$ of the automorphism on the 
{\em bundle\/} \cbb\ of \cb s has finite order.

Finally, in section \ref{s.tf} we exploit these results to formulate a 
conjecture for a general formula (equation \erf{conj1}) for the traces 
of the implementing maps $\Tommuvd$. 
This conjecture suggests a generalization to curves of higher genus -- see 
\erf{conj2} -- which is consistent with factorization. 
The expression \erf{conj2} for the traces constitutes a 
generalization of the Verlinde formula. Surprisingly enough we also observe 
that in all examples that have been checked numerically,
the traces turn out to be integers.

Details of various calculations are deferred to appendices.

\sect{Action of \G\ on affine \lie s and their modules}\label{s.Au}

We start our investigations by introducing a certain family of \auto s of 
affine \lie s which we call \gsauto s. To determine the class of a \gsauto\ 
modulo inner automorphisms, we study its implementation on \ihwm s.

\subsection{\Gsauto s of affine \lie s}

In this subsection we introduce a distinguished class of \auto s of untwisted
affine \lie s \g. Such a \lie\ \g\ can be regarded as a centrally extended 
loop algebra, i.e.\,%
 \futnote{Often it is the sub\alg\ $\oog\eq\gB\otimeS
\complex[t,t^{-1}] \oplus \complex K$ of \g\ that is generated
when one allows only for Laurent polynomials rather than arbitrary Laurent
series that one refers to as the affine \lie. However, in \cft\ one
deals with \g-\rep s for which
every vector of the associated module (\rep\ space) is annihilated by
all but finitely many generators of the sub\alg\ $\oog{}_+$ of $\oog$
that corresponds to the positive roots. Any such \rep\ of $\oog$ can be 
naturally promoted to a \rep\ of the larger \alg\ \g.}
  \be  \gM:= \gB\otimes\complex\twobrac t \oplus \complex K  \,,  \labl g
where \gb\ is a \findim\ simple \lie\ and $t$ an indeterminate.
We identify \gb\ with the zero mode subalgebra of \g, i.e.\ $\gB\ni\xb
\equiv\xb\ot t^0 \in\g$, and refer to it as the {\em horizontal\/} subalgebra
of \g; the rank of \gb\ will be denoted by $r$. We may already remark at this 
point that our main interest in the \gsauto s of the affine \lie\ \g\ stems 
from the fact that collections of $m$ such \auto s combine in a natural way 
to \auto s of $m$ copies of \g\ with identified centers.
The latter automorphisms, in turn, restrict to automorphisms of 
an important class of \infdim\ subalgebras, the so-called \bla s \gT;
those \auto s of \gT\ will be introduced in section \ref{s.au}.
In the present section, however, we study the \auto s of \g\ in their own right.

We will describe the \auto s of the \alg\ \g\
by their action on the canonical central element $K$ and on the elements 
$\hi\ot f$ and $\eal\ot f$ of \g, where $f\iN\complex\twobrac t$ is arbitrary
and $\{\hi\Mid i\eq\onetor\}\cup\{\eal\Mid\alphab\;{\rm a}\;\gb\mbox{-root}\}$
is a Cartan\hy Weyl basis of \gb. (By choosing $f\eq t^n$ with
$n\iN\zet$ one would obtain a (topological) basis of \g, but for the present
purposes it is more convenient to write the formul\ae\ for general 
$f\iN\complex\twobrac t$.)
In terms of these elements, the relations of the \lie\ \g\ read 
$[K,\cdot\,]\eq0$ and
  \be  \bearl
  [\hi\ot f,H^j\ot g] = G^{ij}\,\Res({\rm d}f\,g)\,K \,, \nxl6
  [\hi\ot f,\eal\ot g] = \alphab^i\, \eal  \oT fg \,, \nxl6
  [\eal\ot f,\ebe\ot g] = \e\alphab\betab \, \ealbe \oT fg \qquad\mbox{for
  $\alphab+\betab$ a \gb-root}\,,  \nxl8
  [\eal\ot f,\emal\ot g] = \sumre i\alphab_i\, \hi \oT fg + \killing\,
  \Res(\rmd f\,g)\,K \,.  \eear \Labl0r
Here $\rmd\,{\equiv}\,\partial/\partial t$, and the residue 
$\Res\,{\equiv}\,\res0$ of a formal Laurent series in $t$ is defined by
  \be  \Res\llb\sum_{n=n_0}^\infty c_n\,t^n\lrb = c_{-1}  \,;  \ee
$G^{ij}$ is the symmetrized Cartan matrix of the \hsa\ \gb, and
$\e\alphab\betab$ is the two-cocycle that furnishes structure constants of \gb\
via $[\eal,\ebe]\eq\e\alphab\betab\ealbe$ when $\alphab{+}\betab$ is a \gb-root.

The \auto s to be constructed are characterized by 
a sequence of elements $\mus$, for $s\eq\onetom$ with $m$ an integer $m\gE 2$,
of the {\em coweight lattice\/} \lwv\ of \gb\ (i.e.\ the lattice dual to the 
root lattice of \gb) that add up to zero. The set
  \be  \G:= \{ \muv\,{\equiv}\,(\mub_1,\mub_2\Ldots\mub_m) \,{\mid}\, 
  \mus\iN\lwv\;{\rm for\;all} \;s\eq\onetom;\;\mbox{$\summs$}\mus\eq0\}  \labl G
of such sequences forms a free abelian group \wrt elementwise addition.
As an abstract group, we have the isomorphism
  \be  \G \cong (\lwv)^{m-1}  \,.  \ee
In addition, the \auto s of our interest also depend on
a sequence of pairwise distinct complex numbers $z_s$ ($s\eq\onetom$),
one of which is singled out. However, in contrast to the elements of \G,
in this section we regard these numbers as fixed once and for all,
and accordingly we do not refer to them in our notation for the \auto s.
Thus we will deal with precisely one \auto\ of \g\ for each $\muv\iN\G$; we 
denote this map by $\ommuv$.
Only in case we wish to stress the dependence of the \auto\ on the label 
$\so\iN\{\onetom\}$ that is singled out, we employ instead the notation 
$\ommuvo$.

For any $\muv\iN\G$,
the \auto\ $\ommuv\equiv\ommuvo$ is defined by 
  \be  \bearl  \ommuv(K) := K \,, \Nxl11 \ommuv(\hi\ot f):= \hi\ot f 
  + K \, \summsp \musp^i\, \Res(\fiss\,f) \,, \nxl9
  \ommuv(\ebe\ot f):= \ebe\ot f \cdot \displaystyle\prod_{s=1}^m
  (\fiss)_{}^{-(\musp,\betab)}  \eear \Labl Au
with
  \be  \fiss(t) := \llb t+(z_\so\mi z_s)\lrb_{}^{-1} \,.  \labl{fiss}
Here for $q\iN\dl Q$ we employ the notation $f^q$ for the function with values
  \be  f^q(z) = (f(z))^q  \Labl5c
for all $z\iN\complex$. In the sequel we will also use the short-hand
  \be  \prd\alphab := \prod_{s=1}^m(\fiss)_{}^{-(\mus,\alphab)}  \labl{prd}
for $\alphab$ a \gb-root.

It is readily checked that the mapping \Erf Au indeed constitutes an \auto\ of
the affine \lie\ \g. Indeed, except for the bracket $[\eal\ot f,\emal\ot g]$,
all the relations \Erf0r are trivially left invariant. As for the latter,
invariance follows by the identity
  \be  \bearll \Res\llb {\rm d}[f\Cdot\prd\alphab]
  \cdot g\Cdot\prd{-\alphab} \lrb\, K  \!\!
  &= \Res\llb{\rm d}f\,g +fg\,{\cdot}\summsp(\mus,\alphab)\,\fiss \lrb\, K \nxl8
  &= \Res({\rm d}f\,g)\,K + \sumre i \alphab_i\, \summsp \mus^i\,
  \Res(\fiss\,fg)\, K \,,  \eear  \Labl1u
which follows with the help of
  \be  \rmd\prd\alphab = \prd\alphab \Cdot \summs(\mus,\alphab)\,\fiss \,.
  \Labl1f

Moreover, it follows immediately from the definition that
  \be  \sigma^{}_{\vec{\Bar\mu}'}\circ\ommuv
  = \sigma^{}_{\vec{\Bar\mu}+\vec{\Bar\mu}'}
  = \ommuv\circ\sigma^{}_{\vec{\Bar\mu}'} \,.  \ee
Thus the \auto s \Erf Au respect the group law in \G, i.e.\ they
furnish a \rep\ of the abelian group \G\ in \autg. This \rep\ 
provides in fact a group {\em iso\/}morphism; 
accordingly we will henceforth
identify the group of \auto s $\ommuv$ \Erf Au with the group \G\ \erf G.

As already mentioned, we
refer to the \auto s \Erf Au as {\em\gsauto s\/} of the affine \lie\ 
\g. The origin of this terminology is that they are close relatives of the 
ordinary \sauto s $\ommuo$ of \g\ that appear in the literature, e.g.\ in 
\cite{frha,goom,levw}.
In the conventional notation 
  \be  H^i_n\df \hi\ot t^n \,, \qquad E^\alphab_n\df\eal\ot t^n \,,  \Labl3h
the action of such a `single-shift' \auto\ $\ommuo$ is given by
  \be  \bearl  \ommuo(K) = K \,, \nxl6 \ommuo(H^i_n) = H^i_n
  + \mub^i\, \delta_{n,0} \, K \,, \nxl4
  \ommuo(E^\alphab_n) = E^\alphab_{n+(\mub,\alphab)} \,,  \eear \Labl A0
where $\mub\iN\lwv$.
These ordinary \sauto s are recovered from the formula \Erf Au as
the special case where $\musp\eq0$ for $s\nE\so$. In the case of \Erf A0,
the coweight lattice element $\mub$ is known as a {\em shift vector\/}; 
we will use this term also for the collection $\muv$ of
coweight lattice elements that characterizes the \gsauto s \Erf Au.

Note that the ordinary \sauto s do not satisfy the constraint
$\summs\mus\eq0$, i.e.\ do not belong to the class of \auto s of \g\ that
we consider here. In this context we should point out that when we dispense of
the restriction $\summs\mus\eq0$, the formula \Erf Au still provides us with
an \auto\ of \g. The reason why we nevertheless impose that constraint 
will become clear in the next section, where we glue together several of
the `local' objects \Erf Au to a `global' object that only exists when
the constraint is satisfied. 

\subsection{Implementation on modules}\label{ss.imp}

A natural question is to which class of outer modulo inner \auto s a \gsauto\
$\ommuv$ belongs. Before we can address this issue, we first have to study the
implementation of $\ommuv$ on \ihwm s.

We start by observing that for any \auto\ $\sigma$ of an 
affine \lie\ \g\ and any irreducible highest weight module $(R,\hil)$ over \g,
  \be  (\tilde R,\hil) := (R\Circ\sigma,\hil)  \ee
furnishes again an irreducible highest weight module.
Since this fact is interesting in itself, we pause to describe how it can be
established. First, irreducibility of $(\tilde R,\hil)$ is immediate: if a 
subspace $M$ of $\hil$ is a non-trivial submodule under $\tilde R$, 
then it is a non-trivial submodule under $R$ as well; but by the
irreducibility of \hil\ such submodules do not exist.
To check the highest weight property, it is sufficient to show that the module
$(\tilde R,\hil)$ possesses
at least one singular vector; this vector is necessarily unique
(up to a scalar), since otherwise the module would not be irreducible.
Now consider any Borel subalgebra \bp\ of \g; then
$\sigma(\bp)$ is a Borel subalgebra again. Since
any two Borel subalgebras are conjugated under an inner automorphism of \g\
\cite{peka}, it follows that there exists an inner automorphism $\sigmat$ of \g\
such that $\sigma(\bp)\eq\sigmat(\bp)$. This implies that
the two automorphisms $\sigma$
and $\sigmat$ just differ by an automorphism that does not affect the
triangular decomposition, i.e.\ by a so-called {\em diagram automorphism\/}
$\omega$ (see e.g.\ \cite{fusS3}). Let us write
  \be  \sigmat = \prod_i^\to \exp(\ad{x_i}) \,,  \ee
where a fixed ordering of the product is understood. 
We consider the vector
  \be  \tilde v:= \prod_i^\leftarrow \exp(R(x_i))\, v \,, \ee
where the opposite ordering of the product is understood 
and where $v$ is the highest weight vector of $(R,\hil)$. 
(One can think of $\tilde v$ as a Bogoliubov transform of $v$.) 
One can check that 
  \be  R(\sigma(x))\,\tilde v = R(\sigmat\omega(x))\, \tilde v
  = \prod_i^\to \exp(R(x_i))R(\omega(x))\, v  \Labl3x
(compare also the relation \Erf3r below).
This tells us that the highest weight properties of $v$ imply analogous
highest weight properties for $\tilde v$. This concludes the argument.

In addition, by similar arguments one can show that when $\sigma$ is
an inner \auto, then the \g-modules $(R,\hil)$ and $(\tilde R,\hil)$ are
isomorphic, and hence share the same highest weight. On the other hand, they do
not necessarily share the same highest weight vector, of course. Thus in 
general the
weight of the \hwv\ $v$ of $(R,\hil)$ is different from its weight as a vector 
in $(\tilde R,\hil)$; but the previous statement tells us that the difference
must be an element of the root lattice of \g.

We also note that any diagram automorphism $\omega$ restricts to an \auto\
of the \csa\ \go\ of \g\ and hence by duality induces an \auto\ $\omd$
of the weight space of \g, which constitutes a permutation 
$\Lambda\,{\mapsto}\,\omd\Lambda$ of the finitely many integrable weights at 
fixed level. It follows that for each integrable \g-weight 
$\Lambda$ any \gsauto\ $\ommuv$ of \g\ induces a map 
  \be  \tommuv\equiv\tommuvo:\quad
  \hil_\Lambda \to \hil_{\omd\Lambda}  \labl{tommuv}
between the \ihwm s \hl\ and $\hil_{\omd\Lambda}$ over \g\
which obeys the twisted intertwining property
  \be  \tommuv\,x = \ommuv(x)\,\tommuv  \Labl Tx
for all $x\iN\gM$; the map $\tommuv$ is characterized by this property
up to a scalar factor.
In the special case where $\ommuv$ is an inner \auto\ of \g\ and hence of
the form $\ommuv\,{=}\stackrel{\to\;}{\prod_i}\exp(\ad{x_i})$, 
the map $\tommuv$ can be implemented by
  \be  \ttommuv = \prod_i^\to\exp(x_i) \,, \Labl tx
where the \rhs\ is to be understood as an element of the universal enveloping
\alg\ of \g. Notice, though, that the implementation $\tommuv$ is only 
determined up to a phase. Indeed, owing to the fact that $K$ is central so that
$\expad{\xi K}\eq\id$, we can always modify $\tommuv$ by central terms,
  \be  \tommuv = \exp(f\,K)\prod_i^\to\exp(x_i) \,,  \Labl t'
where $f$ is an arbitrary element of \fun. We will make use of this freedom
later on. Moreover, notice that even if all automorphisms in 
$\ommuv\eq\prod_i\exp(\ad{x_i})$ commute so that this product is well-defined
without specifying the ordering, this does not imply that the same holds
for the product that appears in the implementation \erf{tx}. 
We will therefore choose some specific ordering of the insertion points and 
keep this ordering fixed in the sequel.

\subsection{Outer automorphism classes}\label{ss.ou}

We are now in a position to determine in which outer automorphism class $\ommuc$
a \gsauto\ $\ommuv$ lies. To this end we take for the Laurent series $f$ the 
constant function \id\ and exploit the 
transformation property of the elements $\hi\ot\id$. 
Because of the identity
  \be  \Res(\fiss\,t^n) = \left\{ \bearll 1 &{\rm for}\ s\eq\so,\;n\eq0 \,,\nxl9
  - (z_s{-} z_\so)^n_{} & {\rm for}\ s\nE\so,\;n\lT0 \,,\nxl9 0 & {\rm else}
  \eear\right.  \Labl1g
for $n\iN\zet$, we have in particular
  \be  \Res(\fiss) = \delta_{s,\so} \,,  \ee
so that 
  \be  \ommuvs(\hi\ot\id) = \hi\ot\id + \mus^i\,K \,.  \Labl1h
Expressed in terms of the notation \Erf3h, the transformation \Erf1h reads 
  \be  H^i_0 \,\mapsto\, H^i_0 + \mus^i\, K \,.  \Labl2h

When combined with the results of subsection \ref{ss.imp}
it follows that modulo the root lattice \lr\ of \gb\
the highest weight $\Lambda$ of any \ihwm\ $(R,\hil)$ over \g\ and
the highest weight $\tilde\Lambda$ of $(\tilde R,\hil)\,{\equiv}\,
(R\Circ\ommuvs,\hil)$ are related by
  \be  \tilde{\Lambdab}-\Lambdab = \kk\,\mus\,\mod\lr \,,  \ee
where $\kk$ is the eigenvalue of the central element $K$. 
Note that this is a statement about the 
{\em horizontal\/} parts of the \hw s; the transformation of the rest of the
weight is then fixed by the fact that because of $\ommuvs(K)\eq K$
the level does not change, while the change in the grade is prescribed by
the formula for $\ommuvs(L_0)$ that is given in \Erf Ln below.\,%
 \futnote{Of course the property that, modulo \lr, for fixed level
there is a universal shift in the horizontal part 
of the \hw\ is not shared by arbitrary \auto s of \g.}

Now the quantity $\lambdab\mod\lr$ is nothing but the {\em conjugacy class\/} 
of the \gb-weight $\lambdab$.  Taking also into account the relation
$\kk\eq\kv\Cdot(\Bar\theta,\Bar\theta)/2$ (with $\Bar\theta$ the highest 
root of \gb) between the eigenvalue $\kk$ of $K$ and
the level $\kv$, we learn that the \auto\ $\ommuvs$ leads to a change $\delta_s$
in the conjugacy class of level-1 modules that is given by 
$(\Bar\theta,\Bar\theta)/2$ times the shift vector $\mus$, taken modulo the
root lattice \lr, or what is the same,
  \be  \delta_s = \Frac{(\Bar\theta,\Bar\theta)}2\, \llb \mus \mod \lv \!\lrb
  \,.  \ee

Furthermore, there is a
natural one-to-one correspondence between conjugacy classes of
\gb-weights and classes of certain outer automorphisms of \g.
Namely, as abelian groups we have the isomorphisms
  \be  \lw/\lr \cong \lwv/\lv \cong {\cal Z}(\g) \,,  \Labl cz
where ${\cal Z}(\g)$ is the unique maximal abelian normal subgroup of the
group $\outg\eq\autg/\intg$ of outer automorphism classes of \g. 
It follows in particular that $\ommucs$ is the same outer automorphism class 
as the class $[\ommuos]$ of the ordinary shift \auto\ $\ommuo$ (compare formula 
\Erf A0) that is characterized by the same shift vector $\mub\eq\mus$.
It is important to note that in general this way one cannot obtain {\em all\/}
outer automorphism classes of \g, but rather only those which belong to the
subgroup ${\cal Z}(\g)$.

Let us also mention that 
the abelian subgroup ${\cal Z}(\g)$ of the group of diagram automorphisms
is naturally isomorphic to the center of 
the compact, connected and simply connected real Lie group whose \lie\ is
the compact real form of \gb. Furthermore, the elements of ${\cal Z}(\g)$ 
are in one-to-one correspondence with the simple currents \cite{scya,intr}
of the \wzwt,
which are those primary fields which correspond to the units of the fusion 
ring of the theory (see e.g.\ \cite{scya6}).

\subsection{Extension to the Virasoro \alg}

Given a \rep\ of the untwisted affine \lie\ \g, the affine
Sugawara construction provides us with a \rep\ of the Virasoro algebra $\vir$.
Recall that \vir\ is the \lie\ with generators $L_n$ for $n\iN\zet$ and $C$
and relations
  \be  [L_n,L_m] = (n-m)\,L_{n+m} + \Frac1{24}\,(n^3-n)\, \delta_{n+m,0}\,C \,,
  \qquad [C,L_n]=0 \,.  \ee
The algebra $\vir$
combines with \g\ as a semidirect sum $\vir\splus\g$, with relations
  \be  [L_n,\xb\ot f] = -\xb\oT t^{n+1} \rmdd f  \Labl lx
for $n\iN\zet$.

As we show in appendix \ref{AV}, the \gsauto s $\ommuvo$
of \g\ can be extended in a unique manner to \auto s of $\vir\splus\g$. The 
action of such an \auto\ on the generators of \vir\ reads $\ommuvo(C)\eq C$ and
  \be  \ommuvo(L_n) = L_n + \sumlz \sum_s (\mus{,}H_{n-\ell})\,
  \Res(t^\ell\fiss) + \Frac12\,K\sum_{s,s'}(\mus{,}\musP)\,\Res(t^{n+1}\fiss
  \fisP)  \Labl Ln
(since the extension is unique, we employ the
same symbol for the \auto\ of $\vir\splus\g$ as for the \auto\ of \g).
Note that by putting $\musp\eq0$ for $s\nE\so$, this reduces to
  \be  \ommuo(L_n) 
  = L_n + (\mub,H_n) + \Frac12\,K\,(\mub,\mub)\,\delta_{n,0} \,;  \Labl0L
this is precisely the unique
extension to the \vira\ of the ordinary \sauto\ \Erf A0 of \g.

For later reference we also mention that via the formula
\Erf Tx and the affine Sugawara construction, the maps
$\tommuv{:}\;\hil_\Lambda \to \hil_{\omd\Lambda}$ defined in \erf{tommuv}
obey the twisted intertwining relations
  \be \tommuvo L_n = \ommuvo(L_n)\, \tommuvo  \Labl TX
with regard to the \vira.

\sect{Action of \G\ on the \bla}\label{s.au}

The \auto s constructed in section \ref{s.Au} also provide us with 
corresponding \auto s of direct sums of affine \lie s with identified
centers. In this section we show that such an \auto\ preserves a 
distinguished subalgebra of the latter \alg, namely the so-called \bla\ which 
plays a crucial role in the description of \cb s.

\subsection{The block \alg}

In the previous section we have established that
given a shift vector $\muv\iN\G$ together with a definite choice of
distinguished label $\so\iN\{\onetom\}$, the prescription
\Erf Au provides us with an \auto\ of the affine \lie\ \g.
But the formula \Erf Au of course makes sense for every choice $\so\eq\onetom$
of this distinguished label, so that we are in fact
even given a collection $\{\ommuvo\}$ of such \auto s, one for each value of 
$\so$, and all with the same shift vector $\muv\iN\G$. By a slight
change of perspective
we can rephrase this by saying that we are given an \auto\ of the direct sum 
$\bigoplus_{s=1}^m\!\gs$ of $m$ copies $\gs\,{\cong}\,\g$, namely the one 
that on the $s$th summand $\gs$ acts as $\ommuvs$. Moreover, the so obtained 
\auto\ restricts to an \auto\ of the quotient
  \be  \gM^m := \llb \bigoplus_{s=1}^m\gs \lrb \mbox{\Large$/$}\!
  \raisebox{-.23em}{$\cal J$} \,, \quad\
  {\cal J}:= \langle K_s{-}K_{s'}\,|\, s,s'\eq 1,2,...,m \rangle  \Labl gM
of $\bigoplus_{s=1}^m\!\gs$ that is obtained by identifying the centers of the 
\alg s $\gs$, because the ideal $\cal J$
is $\ommuv$-invariant.

While the specific form of the \auto s is rather irrelevant for this simple 
observation, it will become crucial in the considerations that follow now,
where we focus our attention to a different \infdim\ \lie\ that can be
embedded into \Erf gM. A convenient starting point for the description of
this \alg\ is to recall that the \auto s \Erf Au also depend on a chosen 
sequence of $m$ pairwise distinct numbers $z_s\iN\complex$.
We now regard these numbers as the coordinates of $m$
pairwise distinct points $p_s$ on the complex projective line \pe, to which
we refer as the {\em insertion points\/} or {\em punctures\/}.
More precisely, we regard the complex curve
\pe\ as $\complex\cup\{\infty\}$ and
denote by $z$ the standard global coordinate on \Complex; then we write
  \be  z_s = z(p_s) \,.  \ee
For the present purposes the points $p_s$ are kept fixed; later on we
will allow them to vary over the whole moduli space $\calm_m$ of 
$m$ pairwise distinct points on \pe.

Given this collection of points, we can consider the space \fun\ 
of algebraic functions on the punctured Riemann sphere 
${\dl P}^1_\mm\,{\equiv}\,\pe{\setminus}\{p_1,p_2\Ldots p_m\}$,
i.e.\ of meromorphic functions on \pe\ whose poles are of finite order and
lie at most at the
punctures $\{p_1,p_2\Ldots p_m\}$. The space \fun\ is an associative \alg, 
the product being given by pointwise multiplication. As a consequence,
the tensor product 
  \be  \gT \equiv \gb \otimes \fun  \Labl gt
of the horizontal subalgebra \gb\ of \g\ with the space \fun\
inherits a natural \lie\ structure from the one of \gb, with Lie bracket
  \be  [\xb\ot f,\yb\ot g] = [\xb,\yb] \oT fg  \Labl xy
for $\xb,\yb\iN\gb$ and $f,g\iN\fun$. In terms of the \gb-generators $\hi$
and $\eal$, this yields the same relations as in \Erf0r, but with the residue
terms removed, and with $f,g$ elements of \fun\ rather than formal power series.

The algebra \gT\ provides a global realization of the symmetries of a WZW
\cft, which are locally realized through the affine \lie\ \g.
As will be described in more detail later on, it constitutes
an important ingredient in a \rep\ theoretic description of the
chiral blocks of \wzwts;
for brevity, we will therefore refer to \gT\ as the {\em\bla\/}. 
Clearly, the \bla\ is spanned by the elements $\hi\ot f$ and $\eal\ot f$ with 
$i\iN\{\onetor\}$, $\alphab$ a \gb-root and $f\iN\fun$.
(By allowing for arbitrary $f\iN\fun$ of course
we do not obtain a basis of the \bla; while a basis can easily be given, we
will not need it here. Note that an \auto\ of \gT\
is uniquely defined by its action on a basis, and
hence a fortiori by its action on the elements $\hi\ot f$ and $\ebe\ot f$.)

Around each of the punctures $p_s$ we can choose the local coordinate 
  \be  \z s= z\mi z_s \,.  \Labl zs
By identifying these local coordinates with the indeterminate $t$ of the loop 
construction of the affine \lie\ $\gs\,{\cong}\,\g$, one obtains an embedding 
of the \bla\ \gT\ in the direct sum of $m$ copies of the
loop algebra $\gloopb\,{\cong}\,\gB\otimeS\complex\twobrac t$, and thereby 
also an embedding $\imath$ in the algebra $\gM^m$ that was introduced in 
\Erf gM. This is seen as follows.
For any function $f\iN\fun$ we denote its expansion in local coordinates 
around the point $p_s$, considered as a Laurent series in the variable 
$t\equiv\z s$, by $f\ssn s$. With this notation, 
the image of $\xb\ot f\iN\gt$ under the embedding $\imath$ is the sequence
  \be  \imath(\xb\ot f) = (\xb\ot f\ssn1,\xb\ot f\ssn2\Ldots \xb\ot f\ssn m)
  \,,  \ee
where $\xb\ot f\ssn s$ is regarded as an element of $\gs$. 
In short, the embedding in the $s$th summand is obtained by replacing the
global function $f$ by its local expansion $f \ssn s$ at the puncture $p_s$,
or in other words, by its germ at $p_s$. Note that $f$ is already determined
completely by its germ at any single puncture. In particular, the block
\alg\ is a proper sub\alg\ of $\gM^m$.

\subsection{Automorphisms of the \bla}

Next we observe that also the functions $\fiss$ defined in \erf{fiss} are
the germs of globally defined functions in \fun. Indeed, in view of \Erf zs
for each pair $\so,s$ the function $\fiss$ 
\erf{fiss} can be recognized as the local expansion
  \be  \fiss(\z\so) = (\z\so+z_\so\mi z_s)_{}^{-1} = \fss(\z\so)  \ee
at $p_\so$ of the function $\feSp\iN\fun$ defined by
  \be  \feS(z) := (z-z_s)_{}^{-1} \,.  \labl{ge0}
At this point the meaning of
the requirement $\summs\mus\eq0$ becomes clear: it ensures that the function
$\prod_s(\feS)_{}^{-(\mus,\betab)}$ possesses
poles only at the punctures $p_s$ and hence lies in \fun.
It therefore follows in particular that the functions
$f\Cdot\prod_s(\fiss)_{}^{-(\mus,\betab)}$ that appear in the \auto\ 
\Erf Au of \g\ are the local germs at $p_\so$ of globally defined functions
$f\Cdot\prod_s(\feS)_{}^{-(\mus,\betab)}$. 
Of course, each of the germs individually already contains all information
on the global function; this is reflected by the fact that for all choices
of $\so$ the Laurent series $(\fiss)_{}^{-(\mus,\betab)}$ all involve one and
the same shift vector $\muv$.

{}From these observations we finally deduce that
to any shift vector $\muv\iN\G$ we can associate a linear
map of the \bla\ \gT\ which acts as
  \be  \Ommuv(\hi\ot f) = \hi\ot f \,, \qquad
  \Ommuv(\ebe\ot f) = \ebe\ot f \cdot \prod_{s=1}^m (\feS)_{}^{-(\mus,\betab)}
  \,.  \Labl aU
It is readily checked that this map constitutes an \auto\ of \gT.
Note that on purpose here we 
employ a similar symbol $\Ommuv$ as for the \auto s $\ommuv$ of \g\
in \Erf Au; indeed these maps should be regarded as the global and
local realizations, \resp, of one and the same basic structure.
To be precise, the local expansions of the \auto\
\Erf aU of the \bla\ reproduce the \auto s \Erf Au of $\gs$ only up to
the central terms. The latter are needed in order to really have an \auto\
of $\gs$ rather than only an \auto\ of the corresponding loop \alg; however,
because of the identification of the centers of the subalgebras $\gs$ in
$\gM^m$ \Erf gM, upon summing over all insertion points these terms
cancel owing to the residue theorem: 
  \be \bearll  \Ommuv(\hi\ot f) \!\!
  & \hat=\; \Ommuv\llb \summs H^i_{\sss(s)}\ot f\ssn s\lrb
   = \summsp \ommuvs(\hi\ot f\ssn s) 
     \;\hat=\; \hi\ot f +K \dsum_{s,s'=1}^m \mub^i_{s'}\,\Res(\fisp\,f\ssn s)
  \Nxl11
  &= \hi\ot f + K \dsum_{s'=1}^m \mub^i_{s'} \cdot \summs \Res\llb (\feSP f)
   \ssn s\lrb = \hi\ot f \,.  \eear \Labl1t

As the reader may already have noticed, for any \gb-root $\alphab$
not only the poles, but also the zeroes of the meromorphic function 
$\prod_s(\feS)_{}^{-(\mus,\alphab)}$ occur at the punctures $p_s$.
In other words, these functions belong to the subset
  \be  \funz := \{ f\iN\fun \Mid f^{-1}\iN\fun \}  \labl{funz}
of \fun, where in accordance with the prescription \Erf5c the symbol $f^{-1}$ 
stands for the function that has values inverse to those of $f$, i.e.\
$f^{-1}(p)\eq(f(p))^{-1}$ for all $p\iN\pe$.
The elements of this subset $\funz\,{\subset}\,\fun$, which is in fact a 
subalgebra, are called the invertible elements or {\em units\/} of \fun.

At this point it is appropriate to mention that the point at infinity of 
$\pe\,{\cong}\,\complex\,{\cup}\{\infty\}$ is by no means distinguished 
geometrically, but acquires its special role only through the choice of
coordinates. Choosing a different quasi-global coordinate $\tilde z$ on \pe\ 
one would assign the coordinate value $\infty$ to a different geometrical point.
Accordingly we should also allow for $z\eq\infty$ as the value of $z$ at
one of the insertion points, for definiteness say $z_m\eq\infty$.
As it turns out, in this situation the discussion above indeed goes through,
but with some modifications. 

An obvious modification consists in
the fact that in place of $\z s\eq z\mi z_s$ \Erf zs a good local
coordinate at $z_m\eq\infty$ is provided by $\z m\eq z^{-1}$.
Accordingly we must e.g.\ replace the formula \erf{fiss} for $\fiss$ by
  \be  \fiss(t) := \left\{ \bearll 
  (t+z_\so\mi z_s)_{}^{-1} & {\rm for}\ z_\so\nE\infty\,,\\[.5em]
  (t^{-1}\mi z_s)_{}^{-1}     &{\rm for}\ z_\so\eq\infty\,.
  \eear \right. \labl{fiss'}
However, this is not the only change that needs to be made. One finds that in
fact in the formula \Erf Au for the \auto s of \g\ one must now only include 
the contributions from $s\iN\{\onetome\}$, but not from $s\eq m$, i.e.\ one now
has
  \be  \ommuvo(\hi\ot f):= \hi\ot f + K \summse \mus^i\,\Res(\fiss\,f) \,,
  \qquad \ommuvo(\ebe\ot f):= \ebe\ot f \Cdot \prd\betab \,. \Labl2u
where the definition \erf{prd} of $\prd\alphab$ is to be replaced by
  \be  \prd\alphab := \prod_{s=1}^\me(\fiss)_{}^{-(\mus,\alphab)} \,.
  \labl{prdi}
Moreover, the formula \Erf2u applies only for $\so\iN\{\onetome\}$, while for
$\so\eq m$ it gets replaced by
  \be  \ommuvm(\hi\ot f):= \hi\ot f - K \summse\mus^i\,\Res(t^{-2}\fiss\,f) \,,
  \qquad \ommuvm(\ebe\ot f):= \ebe\ot f \Cdot \prd\betab \,. \Labl3u
Similarly, instead by \Erf aU, the associated \auto\ of the \bla\ \gT\ is now
given by 
  \be  \Ommuv(\hi\ot f) = \hi\ot f \,, \qquad
  \Ommuv(\ebe\ot f) = \ebe\ot f \cdot \prod_{s=1}^\me (\feS)_{}^{-(\mus,\betab)}
  \,.  \Labl au

After these replacements all calculations go through as before. 
In particular, still the functions $\prod_{s=1}^\me(\fiss)_{}^{-(\mus,\betab)}$
are the local germs of the globally defined function
$\prod_{s=1}^\me(\feS)_{}^{-(\mus,\betab)}$ at $p_\so$ for {\em all\/}
$\so\iN\{\onetom\}$. Also, that function is still an element of \funz.
(On the other hand, of course now one can no longer invoke the identity
$\summs\mus\eq0$ to exclude poles at infinity; but this is also no longer
needed, since $\infty$ is a puncture.
Explicitly, now the divisor of the function $\prod_s(\feS)_{}^{-(\mus,\betab)}$
consists of the points $p_s$, $s\eq\onetome$, with finite value of $z_s$,
which have multiplicity $-(\mus,\betab)$, and in addition of $\infty$ which 
has multiplicity $\summse(\mus,\betab)\eq{-}(\mub_m,\betab)$.)
Briefly, the function $\prod_{s=1}^\me(\feS)_{}^{-(\mus,\betab)}$ has
already by itself a pole respectively zero of the correct order at $\infty$.

Let us also note that at $z_m\eq\infty$ the formula \Erf1h for
$\ommuvo(\hi\ot\id)$ still applies. Namely, in this case we have
  \be  \ommuvm(\hi\ot\id) = \hi\ot\id - K \, \summse \mus^i
  \, \Res( t^{-2}\,(t^{-1}-z_s)^{-1} ) \,.  \ee
Using the fact that $\Res(t^{-2}(t^{-1}-z_s)^{-1} )\eq1$
independently of $s$ together with the condition that the shift vectors add 
up to zero, this is nothing but
  \be  \ommuvm(\hi\ot\id) = \hi\ot\id + \mub_m^i\,K \,.  \ee
It follows in particular that the statement that the
outer automorphism class $\ommucs$ of $\ommuvs$ can be read off the 
coweight lattice element $\mus$ is still true for $s\eq m$ with $z_m\eq\infty$.
We also mention that in the case $z_m\eq\infty$ the residue theorem still 
ensures the cancellation of central terms as in formula \Erf1t. Namely,
while the summation over $s'$ in \Erf1t is now only from 1 to $\me$, 
the summation over $s$ still ranges from 1 to $m$; the desired result then
follows after taking into account that
$\Res\llb(\feSP f)\ssn m\lrb = \res\infty(\feSP f) = -\res0(t^{-2}\feSP f)$.

Finally we remark that the Sugawara construction provides us at each of
the insertion points $p_s$, with a Virasoro algebra $\vir\,{\equiv}\,\vir_s$. 
We write $L_n^\ssss$ for the generators of the \vira\ $\vir_s$ associated to 
\gS. Also note that in the formula \Erf Ln for the action of a \gsauto\ on
\vir\ the sums over $s$ and $s'$ extend over all punctures except, when 
present, the one at $\infty$.  

\sect{Inner automorphisms}\label{s.II}
 
In subsection \ref{ss.imp} we have seen how an \auto\ of the affine \lie\
\g\ can be implemented on \g-modules. We can now implement the tensor
product of $m$ \gsauto s on the tensor product of $m$ \g-modules.
We will see that this leads to a well-defined map on the spaces of
\cb s because such a tensor product of \gsauto s leaves the 
\bla\ \gT\ invariant. One convenient property that one would like
to achieve for the maps on the \cb s is that they are of finite order.
As it turns out, to this end we must find a suitable description for those
\gsauto s which are inner automorphisms of \gT.

The subset
  \be  \GO:= \{ \muv\,{\equiv}\,(\mub_1,\mub_2\Ldots\mub_m) \,{\mid}\, 
  \mus\iN\lv\;{\rm for\;all} \;s\eq\onetom;\;\mbox{$\summs$}\mus\eq0\}  \Labl GO
of those elements of \G\ for which
the allowed range of the shift vectors $\mus$ is restricted to the
co{\em root\/} lattice $\lv\,{\subseteq}\,\lwv$ of \gb\ clearly is a subgroup 
of \G. In this section we show that the \auto s \Erf Au of \g\ and 
\Erf aU of \gT\ associated to shift vectors that lie in \GO\ are {\em inner\/} 
\auto s of the block algebra \gT, i.e.\ \auto s that can be written 
as $\prod_i\expad{x_i}$ with suitable elements $x_i$ of \g\ \resp\ \gT;\,%
 \futnote{In the present section we only show that $\muv\iN\GO$ is sufficient 
for having an inner \auto. But as we shall see later on, this is also necessary.}
a fortiori they are then also inner automorphisms of the
ambient algebra $\gM^m\,{\supset}\,\gt$. 

For establishing this result we need to introduce some special inner \auto s
of the \bla\ \gT, and for these we need some particular elements of \gT.
For convenience, and \wolog, in this section on we assume that 
0 and $\infty$ are among the insertion points $p_s$, and number the latter
in such a way that $z_m\eq\infty$; this has the technical advantage that
for each insertion point $p_s$ with $s\iN\{\onetome\}$\,%
 \futnote{Recall that there is no function $\feS$ associated to $\infty$.}
the function $\feS$ defined by \erf{ge0}
already lies in the subalgebra \funz\ of units of \fun. The relevant special
elements of \gT\ are then of the form
  \be  \xaf:= \eal \ot f + \emal \ot f^{-1} \,, \labl{xaf}
where $\eal\iN\gb$ is a step operator of the \hsa\ \gb\ and $f\iN\funz$ is a
unit in \fun.

Furthermore, for any two elements $X,Y\iN\gt$ and any complex number $\xi$,
we denote by $\E XY\xi$ the inner \auto\
  \be  \E XY\xi := \expad{-\xi Y } \circ \expad{\xi X}  \labl E
of the \bla. We are interested in \auto s of the form $\E\xafe\xafz{\ipi/2}$ 
with an arbitrary positive \gb-root $\alphab$ and some suitable functions
$f_1,f_2\iN\funz$. By direct computation we find that
  \be  \E\xafe\xafz{\ipi/2}(\hi\ot f) = \hi\ot f  \labl{omH}
for all $i\eq\onetor$ and
  \be  \E\xafe\xafz{\ipi/2}(\ebe\ot f) = 
  \ebe\oT (f_1/f_2)_{}^{-(\alphav,\betab)}\,f  \labl{omE}
for all \gb-roots $\betab$; here $\alphav\eq2\alphab/(\alphab,\alphab)\iN\lv$ is
the coroot associated to the \gb-root $\alphab$.
(Though straightforward, this calculation is somewhat lengthy; some 
relevant formul\ae\ are collected in appendix \ref{AA}.)

The results \erf{omH} and \erf{omE} imply in particular that
any two \auto s of this form commute. Also,
only the quotient $\fE/\fZ$ appears in \erf{omE}, so that \wolog\ we 
can put $\fZ\eq\id$. It follows in particular that when we compose 
inner \auto s of the form $\E\xafe\xafz{\ipi/2}$ with the special choice
$\fE\eq\feS$ as defined by \erf{ge0} for some $s\iN\{\onetome\}$ and 
$\fZ\eq\id$ and with arbitrary \gb-roots $\alphab$, then we arrive at
\auto s of the form studied previously. Indeed,
  \be  \prod_{s=1}^\me \prod_{i_s=1}^{\ell_s} 
  \E{X_{\alphab_{i_s},\feS}}{X_{\alphab_{i_s}^{},\iD}}{\ipi/2} = \Ommuv \,,
  \labl{ommuv}
where $\ommuv$ is the \gsauto\ of the \bla\ \gT\ that acts as in \Erf au, with 
  \be  \vmub = (\mub_1,\mub_2\Ldots\mub_m) \,, \qquad
  \mus=\sum_{i_s=1}^{\ell_s} \alphab_{i_s}^\Vee \in\lv \quad{\rm for}\
  s\eq\onetome  \labl{vmub}
(and where the ordering in the product is irrelevant).
In short, for any choice of the \gb-roots $\alphab_{i_s}^{}$ we obtain an inner
\auto\ of the form \Erf au with $\mus\iN\lv$. Moreover, by appropriately
choosing these roots we can obtain {\em every\/} \auto\ $\ommuv$ \Erf au for 
which $\muv\iN\GO$. It follows in particular that any \auto\ $\Ommuv$ with
$\muv\iN\GO$ is an {\em inner\/} \auto\ of the \bla\ \gT, as claimed.

When we replace the functions $f$ etc.\ that appeared in the
considerations above, which are elements of \fun, by functions that have the 
same mapping prescription but are regarded 
as Laurent series in the relevant local coordinate $\z s$,
\resp\ as formal Laurent series in the variable $t$, then we can immediately
repeat all steps in the derivation of the formul\ae\ \erf{omH} and
\erf{omE}; thereby for each summand \gS, $s\eq\onetome$, of the algebra 
$\bigoplus_{s=1}^m\gs\supset\gM^m$ we construct a certain class of 
commuting inner \auto s. The action of these \auto s differs from the one of the
\auto s of the \bla\ precisely in that
there arise additional terms that are proportional to the central element 
$K\,{\equiv}\, K_s$ of \gS.
We find that \erf{omH} gets modified to (for some details see appendix \ref{AA})
  \be  \E\xafe\xafz{\ipi/2}(\hi\ot f) = \hi\ot f - \alphaV^i K\,
  \llb \Res(\fE^{-1} \rmdd\fE\, f) - \Res(\fZ^{-1} \rmdd\fZ\, f) \lrb
  \,,  \labl{omH2}
while the formula \erf{omE} applies without any change also to the present 
situation.

We now investigate the case where these Laurent series are precisely the local 
expansions of the functions $\feS$ studied previously. 
Thus for any $s,\so\eq\onetome$ we have to deal with the inner 
\auto\ $\Exsys$ of \gS\ with 
  \be  \xs = \eal \ot \fiss + \emal \ot(\fiss)^{-1} \qquad{\rm and}\qquad
       \ys = (\eal\pl\emal) \oT \id \,,  \ee
where $\fiss(t)$ is the local expansion \erf{fiss} at $p_\so$
of the function $\feSp$ as defined by \erf{ge0}.
By inserting these functions in the general formul\ae\ obtained above,
we learn that the \auto\ $\Exsys$ acts as
  \be \bearl
  \Exsys(\hi\ot\fso) = \hi\ot\fso \, + \alphaV^i K\, \Res(\fiss\fso)\,, \nxl4
  \Exsys(\ebe\ot\fso)= \ebe\ot\fso(\fiss)^{-(\alphav,\betab)} \,.  \eear \Labl29

Next we take the product of such \auto s that is analogous to the product in
\erf{ommuv}. Thereby for each $\so\eq\onetome$
we arrive at \auto s $\ommuvo$ of \gso\ which act as $\ommuvo(K)\eq K$ and
  \be  \bearl
  \ommuvo(\hi\ot\fso) = \hi\oT\fso 
  + K \, \sum_{s=1}^\me \mus^i\, \Res(\fiss\,\fso) \,, \nxl6
  \ommuvo(\ebe\ot\fso) = \ebe\oT\fso\Cdot\prod_{s=1}^\me
  (\fiss)_{}^{-(\mus,\betab)} \,.  \eear \Labl0c
For $\so\eq m$ an analogous result holds, with the formula in the first line
replaced by
  \be  \ommuvm(\hi\ot f\ssn m) = \hi\oT f\ssn m
  - K \, \sum_{s=1}^\me \mus^i\, \Res(t^{-2}\fism\,f\ssn m)  \ee
(and with $\fism$ as given in \erf{fiss'}).
Thus again we have succeeded in constructing all those \gsauto s $\ommuv$
-- this time the \auto s \Erf Au of the affine \lie\ \g\ -- for which the
shift vector $\muv$ lies in the subgroup \GO\ of \G, and hence
can conclude that all such \auto s are {\em inner\/} \auto s of \g.

Of course, the function $\fso(\fiss)^{-(\alphav,\betab)}\eq
\fso(\fss)^{-(\alphav,\betab)}$ is nothing but the
local expansion of the function $f(\feSp)^{-(\alphav,\betab)}$ at $p_s$.
Hence in particular upon summation over $s$
the central terms in the transformation \Erf29 of $\hi\ot f$ 
cancel owing to the residue theorem. Thus the \auto s \Erf0c are
the local realizations of the \auto\ \erf{ommuv} of the \bla.

One may also analyze the analogous inner \gsauto s of the \vira.
This is briefly mentioned at the end of appendix \ref{AA}.

\sect{Implementation on \cb s}\label{s.cb}

\subsection{Implementation on tensor products}\label{s.cb1}

The next step is now to implement a tensor product of \gsauto s on a tensor
product of modules of the affine \lie. This gives rise to a projective action 
of the group \G\ on this tensor product and, of course, also to a dual
action on the algebraic dual of the tensor product. As we will explain, the
space \cbs\ of \cb s can be identified with a subspace in this algebraic dual. 
We will see that the action of \G\ can be restricted to \cbs\ and that this 
action has finite order on the subspace \cbs.

Let us first recall the existence of the maps
$\tommuv\equiv\tommuvo$ \erf{tommuv}. They act on the $\so$th factor
of the tensor product 
  \be  \vh \equiv \vhl = \hil_{\Lambda_1}\otimeS\hil_{\Lambda_2}\otimeS
  \cdots\otimeS\hil_{\Lambda_m}  \Labl vh
of \ihwm s over \g, i.e.\ $\tommuv{:}\;\hil_{\Lambda_\so}\,{\to }\,
\hil_{\omd\Lambda_\so}$, with $\omd$ the permutation of integrable weights that
was described before \erf{tommuv}. The tensor product
  \be  \Tommuv:= \tommuvI1\oT \tommuvI2\oT \cdots  \oT\tommuvI m  \ee
of these maps for all $\so\eq\onetom$ provides us with an
analogous map between tensor products,
  \be  \Tommuv:\quad \vhl \to \vhol \,,  \labl{vhol}
where $\omdv\vec\Lambda\eq(\omd\Lambda_1,\omd\Lambda_2\Ldots\omd\Lambda_m)$.
Now note that this map is defined for any $\muv\iN\G$. Moreover, 
  \be  \muv \,\To\, \Tommuv  \Labl To
constitutes a projective \rep\ of \G, i.e.\ 
up to possibly a $U(1)$-valued cocycle $\epsilon$, the group law of \G\ 
is also realized by the action of the maps $\Tommuv$ on the modules,
  \be  \TommuvA\,\TommuvB = \epsilon(\muv_1,\muv_2)\, \TommuvAB \,;  \Labl BA
moreover, the cocycle on \G\ is uniquely determined by a cocycle on the finite
group $\GG/\GGO$.
This result can be established as follows. First we
realize that the \hwv\ of the \g-module \hl\ can be required
to be normalized and then is unique up to a phase. Accordingly the maps
$\tommuvo$ are unique up to a phase, too. Our aim is to show that we can
choose this phase as a function of $\muv$ in a suitable manner.
To this end we proceed in several steps.

We first define the maps $\Tommuv$ for elements $\muv\eq\gammav\iN\GGO$ of
the subgroup \GO\ \Erf GO. This group is a finitely generated free abelian group, so
there is a finite set $\{\betj\}$ of independent 
generators. For instance one can choose these generators to be all of the form
  \be 
  \Bar\gamma_{(j);s_1} = \betab^\Vee \,,\qquad
  \Bar\gamma_{(j);s_2} =-\betab^\Vee \,,\qquad 
  \Bar\gamma_{(j);s} = 0 \quad\ {\rm for}\;\ s\nE s_1,s_2  \Labl7t
with suitable choices of $s_1,s_2\iN\{\onetom\}$ with $s_1\,{<}\,s_2$
and simple \gb-roots $\betab$. Now for each $\betj$ we fix the freedom in
the definition of the implementing maps by explicitly prescribing a preferred 
implementation $\Ttommuvj$ through elements of the block \alg. Concretely,
in the case of generators of the form \Erf7t we set
  \be \bearll  \Ttommuvj := \!\!
  & \id \oT \cdots \oT \id \oT 
  \exp\llb {-}\Frac{\ii\pi}2 X_{\betab,\iD}\lrb   
  \exp\llb  \Frac{\ii\pi}2 X_{\betab,\varphi_{\so,s_1}}\lrb \oT \id \oT \nxl4
  &\hsp{1.68} \cdots \oT \id \oT
  \exp\llb {-}\Frac{\ii\pi}2 X_{\betab,\iD}\lrb   
  \exp\llb  \Frac{\ii\pi}2 X_{\betab,\varphi_{\so,s_2}}\lrb 
  \oT \id \oT \cdots \oT \id \,, \eear \labl{primp}
where the notation $\xaf$ is as introduced in \erf{xaf} and where the 
non-trivial maps occur at the $s_1$th and $s_2$th factor of the tensor product.
A priori it might be expected that the maps $\Ttommuvj$ for different values of 
$j$ do not commute, but rather satisfy
  \be  \Ttommuvj\,\Ttommuvk = \eta(\betj,\betk)\,\Ttommuvj\,\Ttommuvk \,.  
  \Labl TT
But as a matter of fact, when there are non-trivial \cb s in $\vhd$
(which is the case we will be most interested in), then
the numbers $\eta$ introduced this way are equal to
one. To see this, we note that the fact that the implementing maps 
\erf{primp} are realized through elements of the block \alg\
implies (compare the remarks around \Erf3r below) that each of the
$\Ttommuvj$ acts as the identity on \cb s. The result then follows by
acting with both sides of \Erf TT (or more precisely, with their
implementation on the blocks, as defined in formula \Erf Ex below) on a \cb.

We now write any arbitrary $\gammav\iN\GGO$ uniquely as a linear combination
$\gammav\eq\sum_j n_j\,\betj$ and define the image of $\gammav$ under the map 
\Erf To as
  \be  \Tommug:=\prod_j (\Ttommuvj)_{}^{n_j}  \,.  \ee
By the result just obtained, the order of the factors in this product does not 
matter. Also, by the same argument as before one concludes that
  \be  \TommugA\,\TommugB= \TommugAB =\TommugB\,\TommugA \qquad\mbox{for all }\;
  \gammav_1, \gammav_2\iN\GGO \,.  \Labl1s

Further, suppose that we have made a choice also for the implementations
$\Tommuv$ for all other $\muv\iN\GG$ (which we do not yet specify, because the
actual choice is not relevant for the argument). Then again by the same 
reasoning as before, i.e.\ by considering the induced maps on the blocks, 
one deduces that \Erf1s in fact generalizes to
  \be  \Tommug\,\Tommuv = \Tommuv\,\Tommug \qquad\mbox{for all }\;
  \gammav\iN\GGO,\ \muv\iN\GG \,.  \Labl2s

Next we choose some set $\MU\eq\{\muvi\}$ of coset representatives for 
$\GG/\GGO$, which is a finite abelian group. For each of the $\muvi$ we make 
some choice of the implementing map $\Tommuvi$, with the arbitrary phase for the
moment still left unspecified. Having made these choices, every $\muv\iN\G$ 
can be uniquely written as $\muv\eq\gammav(\muv)+\muvim$ with 
$\gammav(\muv)\iN\GGO$ and $\muvim\iN\MU$; we then define
  \be  \Tommuv := \Tommugv\,\Tommugm  \Labl2r
(where the order of the factors is again irrelevant).
Note that this way we have also defined all maps $\Tommuvij$, i.e.\ in 
particular also for those cases where $\muvi{+}\muvj\,{\not\in}\,\MU$. Therefore
we can define phases $\epsilon(\muvi,\muvj)$ by
  \be  \Tommuvi\,\Tommuvj = \epsilon(\muvi,\muvj)\, \Tommuvij \,.  \Labl4s

Finally, combining formula \Erf4s with the result \Erf2s, we learn that for
arbitrary $\muv,\nuv\iN\GG$ we have
  \be  \Tommuv\,\Tonnuv = \Tommugv\,\Tommugm\,\Tomnugv\,\Tommugn
  = \epsilon(\muvim,\muvin)\, \Tommnugv\,\Tommugmn \,.  \ee
When $\muvim{+}\muvin\iN\MU$, this yields immediately
  \be  \Tommuv\,\Tonnuv = \epsilon(\muvim,\muvin)\, \Tomnuv \,,  \ee
while for $\muvim{+}\muvin\,{\not\in}\,\MU$ the same result is obtained after
envoking the definition \Erf2r and the identity \Erf1s.

We thus conclude that, as claimed, the group law of \G\ is realized by the maps
$\Tommuv$ up to a cocycle $\epsilon$, and furthermore this cocycle is
induced from the cocycle on $\GG/\GGO$ that was introduced in formula \Erf4s.
For later reference we also note that (compare e.g.\ \cite{KArp4}) 
the center \CEGC\ of the twisted group \alg\ \CEG\
is the ordinary group \alg\ $\CEGC\eq\complex(\REG)$ of the subgroup
  \be  \REG := \{ [\muvi]\iN\GG/\GGO \,|\,  \epsilon(\muvi,\muvj)\eq
  \epsilon(\muvj,\muvi)\mbox{ for all } [\muvj]\iN\GG/\GGO \} \labl{REG}
of so-called regular elements of $\GG/\GGO$.

At this point it is appropriate to recall that so far we have left the phase 
choices for the maps
$\Tommuvi$ undetermined. Changing these phases will change the cocycle
$\epsilon$ by a coboundary. Thus by adjusting these phases we can achieve to
obtain some preferred representative cocycle in the cohomology class of
$\epsilon$. {}From the cohomological properties of finite abelian groups
(see e.g.\ \cite{BRow'}) it follows in particular that this way we can
achieve the property that all the numbers $\epsilon(\muvi,\muvj)$ are
roots of unity.
In the sequel we will often assume that such a phase choice has been made.

Taken together, these results will allow us to implement the \gsauto s also
on \cb s in such a way that we even obtain a projective action of the 
group \G\ on the space of \cb s. 
To make the implementation of \G\ on blocks explicit, we need an
appropriate description of the \cb s in terms of the action of the \bla.
  
\subsection{Chiral blocks from \coin s}

In a \rep\ theoretic approach, the \cb s of a \cft\ are 
constructed with the help of {\em\coin s\/} $\coi\vh{\liefont B}$ of tensor 
products $\vh \equiv \vhl$ \Erf vh of irreducible modules over the chiral 
algebra \wrt a suitable block \alg\ $\liefont B$ (see e.g.\ \cite{Ueno,fuSc6}).
This statement involves two new ingredients that need to be explained. First,
in general, by a \coin\ of a module $V$ over some \lie\ $\liefont h$ one means
the quotient vector space
  \be  \coi V{\liefont h} := V \,/\, \U^+(\liefont h)V \,,  \Labl3c
where $\U^+(\liefont h)\eq\liefont h\U(\liefont h)$ with
$\U(\liefont h)$ the universal enveloping \alg\ of $\liefont h$.
When the $\liefont h$-module $V$ is fully reducible, then \Erf3c is just the
submodule of $\liefont h$-singlets in $V$, but generically it is a genuine
quotient which cannot be identified with a subspace of $V$. And second, the 
action of the block \alg\ $\liefont B\eq\gt$ on the tensor product vector 
space $\vh$ is defined by its expansions in local coordinates, i.e.\ for 
$X\eq\xb\ot f$ and $v\eq v_1\ot v_2\ot\cdots\ot v_m$ one has
  \be  X\,v := \summs v_1\ot v_2 \ot\cdots\ot v_{s-1}\ot\,(\xb\ot\fs)v_s\,\ot
  v_{s+1}\ot\cdots\ot v_m \,,  \Labl Xv
where $\xb\ot\fs$ is regarded as an element of the loop algebra \gloopb\ and
hence of \g, or more precisely, as the \rep\ matrix of that element of \g\ in 
the \g-\rep\ $R_{\Lambda_s}$. (That this yields a \gT-\rep\ follows by
  \be  \bearll
  \llb (\xb\ot f)(\yb\ot g)\mi(\yb\ot g)(\xb\otim f)\lrb (v_1\ot v_2\ot\cdots\ot
  v_m) \!\!\!
  &= \summsp\!v_1\ot v_2\ot\cdots\ot[\xb\ot f\ss,\yb\ot g\ss]v_s\ot\cdots\ot v_m
  \Nxl12 &\hsp{-4.8}
  = \summsp v_1\ot v_2 \ot\cdots\oT ([\xb,\yb]\ot (fg)\ss)v_s\oT \cdots\ot v_m 
  \Nxl16 &\hsp{-1.9}
  + K\,\kappa(\xb,\yb)\,\llb \summsp\!\res{p_s}(\rmd f\,g)
  \lrb \, v_1\ot v_2 \ot\cdots\ot v_m \,, \eear \labl{xfyg}
where in the first equality one uses the fact that terms acting on different 
tensor factors cancel and in the second equality 
the bracket relations of \g\ are inserted.
The terms in \erf{xfyg} that involve the central element $K$ cancel as a
consequence of the residue formula, while the other terms add up to
$[\xb\ot f,\yb\ot g]\,v$, where the
Lie bracket is the one of \gT\ as defined in \Erf xy.)

The \findim\ vector spaces $\coi\vh{\liefont B}$ of \coin s play the
role of the dual spaces \cbd\ of the \cb\ spaces \cbs, i.e.\
  \be  \cbsl = \llb \coi\vhl{\liefont B} \lrb^\dual_{}  \,.  \ee
By duality, the \cb s\ can then also be regarded as the {\em invariants\/} 
(or in other words, singlet submodules) in the algebraic dual $\vhd$ of $\vh$,
i.e.
  \be  \cbsl = (\vhl^\dual)^{\liefont B}_{} \,;  \ee
in this description, the blocks are \lifo s $\bl$ on $\vh$ with the property
  \be  \pair\bl{Xv} \equiv \bl(Xv) = 0  \Labl bl
for all $X\iN\gt$ and all $v\iN\vh$.

\subsection{Isomorphisms of \cb s}\label{s.gci}

In \Erf3c we followed the habit of suppressing the symbol $R$ for the \rep\ 
by which $\liefont h$ acts on the vector space $V$, e.g.\ $\coi V{\liefont h}$ 
is a shorthand for $\coi V{R(\liefont h)}$. This cannot cause any
confusion as long as we only deal with a single $\liefont h$-module $(R,V)$ 
which is based on the vector space $V$. On the other hand, as is easily
checked, given any \auto\ $\sigma$ of $\liefont h$, together with $(R,V)$ also
$(\tilde R,V)\df (R\Circ\sigma,V)$
furnishes a module over $\liefont h$. Accordingly we are then dealing with
two different actions of $\liefont h$ on $V$, and hence with
two different spaces \Erf3c of \coin s. It is then natural to try to associate 
to any \auto\ $\sigma$ of $\liefont h$ a corresponding
mapping of the respective spaces of \coin s.
This is easily achieved once we are given a linear map $\Tausig{:}\;V\to V$
with the property that $\Tausig\,R(x)\,v\eq\tilde R(x)\,\Tausig\,v$
for all $x\iN\liefont h$ and all $v\iN V$, or in short (suppressing again the
symbol $R$),
  \be  \Tausig\,x = \sigma(x)\,\Tausig  \Labl7a
for all $x\iN\liefont h$.
Namely, we observe that via the prescription $\sigma(\bfe)\df\bfe$ and
linearity, the \auto\ $\sigma$ of $\liefont h$ extends to an \auto\ of the
enveloping \alg\ $\U(\liefont h)$ that respects the filtration,
hence also to an \auto\ of $\Up(\liefont h)$. Because of \Erf7a the 
prescription
  \be  \coi V{R(\liefont h)}\ni [v] \,\mapsto\, [\Tausig v]
  \in\coi V{\tilde R(\liefont h)}  \Labl vv
then supplies us with a well-defined mapping from $\coi V{R(\liefont h)}$
to $\coi V{\tilde R(\liefont h)}$, and by construction this is in fact an 
isomorphism of vector spaces.

Now it is a general fact that {\em inner\/} \auto s act trivially on \coin s.
Namely, an inner \auto\ $\sigma$ of $\liefont h$ can by definition be written as
a product of finitely many \auto s $\sigma_i\eq\expad{x_i}$ for some elements 
$x_i\iN\liefont h$. Moreover, for every $y\iN\liefont h$ we have
(compare the formul\ae\ \Erf3x and \Erf tx)\,%
 \futnote{This may be checked by replacing $x_i$ by $\xi x_i$ and comparing 
both sides order by order in the dummy variable $\xi$.
In the special case that $\liefont h$ integrates to a group, the formula 
is an immediate consequence of the fact that $\exp(\ad x)(y) = \gamma y
\gamma^{-1}$, where $\gamma$ is a group element such that $\gamma\eq\exp(x)$.}
  \be  R\Circ\sigma_i(y) = R(\exp(\ad{x_i})(y)) = \exp(R(x_i))\, R(y)\,
  \exp(-R(x_i)) \,.  \Labl3r
By expanding the exponentials it then follows in particular that
  \be   \tilde R(y)\,v
  = R\Circ\sigma(y)\,v = R(y)\,v \ \bmod\,\Up(\liefont h) V  \ee
for all $v\iN V$ and all $y\iN\liefont h$. By the definition of \coin s, this
means that the modules $(R,V)$ and $(\tilde R,V)$ possess the same \coin s.
By duality, this also means that the dual spaces to these modules possess the
same invariants.

When applied to the situation of our interest, these general observations
tell us that as far as the study of \cb s is concerned we need to regard the
\auto s $\Ommuv$ of the \bla\ \gT\ that we defined in \Erf au above only 
modulo inner \auto s. Accordingly we should determine the outer \auto\ class 
to which a \gsauto\ $\Ommuv$ of \gT\ belongs. We have already addressed this 
question in section \ref{s.Au} for the case of \gsauto s 
of the affine \lie\ \g. Now we study the same issue for the \bla\ \gT. We first
note that the result of section
\ref{s.Au} implies that whenever at least one of the 
vectors $\mus$ is not a coroot, then $\ommuvs$ is an {\em outer\/} \auto\ of the
corresponding affine \lie\ $\gs$; it follows that in this case it is also an 
outer \auto\ of $\gM^m$, and thereby a fortiori
an outer \auto\ of the \bla\ \gT. 

In other words, the subgroup \GO\ \Erf GO
already exhausts the set of those shift vectors $\muv\iN\G$ for which 
$\ommuv$ is an inner \auto\ of \gT.
Thus the group \GQ\ of outer modulo inner \gsauto s of \gT\ is precisely the
factor group 
  \be  \GQ = \G/\GO \,.  \Labl GQ
According to the isomorphism \Erf cz between $\lwv/\lv$ and the unique maximal 
abelian normal subgroup ${\cal Z}(\g)$ of $\outg\eq\autg/\intg$, we thus have
  \be  \GQ \cong ({\cal Z}(\g))^{m-1} \,.  \ee
In particular, \GQ\ is a {\em finite\/} group of order
${\rm ord}(\GQ)\eq|\lw/\lr|^{m-1}$.
For future reference we mention that a set of distinguished representatives of 
the elements of $\outg$ is provided by the diagram \auto s $\omeuvs$ of \g\ 
(see subsection \ref{ss.imp}); thus elements of \GQ\ may be regarded as 
collections of suitable diagram \auto s whose product is the identity.
We also recall that when implementing the group \G\ through
the maps $\Tommuv$ which satisfy \Erf BA, we are effectively dealing with a 
two-cocycle on the finite abelian group $\GG/\GGO\eq\GQ$.

Explicitly, for every $\muv\iN\G$ we have a map $\Tommuvd$ from the \cb s 
$\cbsl\eq(\vhl^\dual)^{\liefont B}$ to $\vhol^\dual$ which acts as
  \be  \pair{\Tommuvd(\bl)}{v} := \pair{\bl}{\TommuvM v} \,.  \Labl Ex
The previous results imply, first, that the image of this map is in \cbsol,
i.e.\ that for any block $\bl$, $\Tommuvd(\bl)$ is again a \cb, and second,
that this map depends on $\muv$ only via its class $[\muv]$ in \GQ. In short,
for each $[\muv]\iN\GQ$ we have constructed an isomorphism
  \be  \Tommuvd \equiv \Tommuvdc:\qquad \cbsl \cong \cbsol \,.  \Labl hh
(Analogously, we also have an isomorphism between the respective dual spaces,
$\coi\vhl\gt \cong \coi\vhol\gt$.)
Note that it already follows from the Verlinde formula and elementary
properties of simple currents that the spaces \cbsl\ and
\cbsol\ have the same dimension; being \findim, it is then trivial
that they are isomorphic as vector spaces. The virtue of the result \Erf hh 
is, however, that it provides us with a canonical realization of this 
isomorphism. As we will soon see, this realization possesses the additional 
non-trivial property to be compatible with a variation of the moduli in the 
problem, i.e.\ with different choices of the insertion points.
Moreover, knowing the isomorphism \Erf hh one can also study the problem
of {\em fixed point resolution\/}, which arises whenever the collection
$\vec\Lambda$ of \g-weights is left invariant by $\omdv$;
this issue will be addressed in section \ref{s.tf} below.

Finally, according to the remarks at the end of subsection \ref{s.cb1}, we can
(and do) employ the freedom in defining the maps $\tommuv$ so as to achieve
the property that all cocycle factors in the projective \rep\ of $\GG/\GGO$
are roots of unity. Together with the fact that $\GG/\GGO$
is a finite group, it follows that the maps $\Tommuvd$ all have finite order.

\sect{Bundles of blocks}\label{s.bu}

In all the considerations above we have regarded the punctures $p_s$ as
held fixed, i.e.\ we have analyzed \bla s and \cb s at a single point of
the moduli space $\calm\equiv \calm_m$ of $m$-punctured projective curves.
We now address the issues that arise when the insertion points are allowed
to vary over the whole moduli space $\calm$. Recall from the discussion in 
subsection \ref{s.cb1}
that in the implementation $\Tommuv$ of an automorphism of $\g^m$ a phase
is still undetermined. Clearly, if we choose this phase at random for each
point in the moduli space, we cannot expect to obtain quantities on the
bundle of \cb s that vary smoothly with the moduli. On the other
hand, the bundle of \cb s carries a natural projectively flat
connection, the \kzc\ \cite{kohn,fesv,math7,Ueno}.
Our rationale will therefore be to implement the
maps $\ommuv(\vz)$ for different values of the moduli $\vz$ in
such a way that they preserve the \kzc.

\subsection{Connections and bundles over the moduli space}

The moduli space $\calm$ is $(\pe)^m$ minus the union of diagonals, i.e.
  \be  \calm = \{ (p_1,p_2\Ldots p_m) \,|\, p_s\iN\pe,\;p_s\nE p_{s'}\;{\rm for}
  \; s\nE s' \} \,.  \ee
When considered as depending on the values of the punctures, the spaces \cbs\
of \cb s combine to a vector bundle \cbb\ over $\calm$ \cite{Ueno}. Before 
we study that bundle, 
we introduce a few other bundles which in this context are of interest as well.

We first consider the trivial bundle over $\calm$ with fiber given by the
\infdim\ \lie\ $\gM^m$ defined in \Erf gM,
  \be  \gM^m \Times \calm \to \calm \, . \labl6b
This bundle carries a fiberwise action of the group \G,
i.e.\ an action by \auto s of the total space that cover the identity on the 
base space $\calm$.
For every $\muv\iN\G$ we write $\ommuv(\vz)$ for the map on the fiber over the
point $\vp\iN\calm$ with coordinates $\vz\,{\equiv}\,(z_1,z_2\Ldots z_m)$.
On the bundle \Erf6b we have a flat connection $D$ which
acts on smooth sections $g(\vz)$ of \Erf6b as
  \be  D_s g(\vz):= \partial_s g(\vz) +  [L_{-1}^\ssss, g(\vz)]  \Labl6c
for $s\eq\onetom$, where $\partial_s\,{\equiv}\,\partial/\partial z_s$ and
$L_{-1}^\ssss\iN\vir_s$ is a generator of the \vira\ associated to \gS.

Moreover, the bundle \Erf6b possesses a {\em subbundle of \bla s\/}, which is 
defined by taking for each $\vp\iN\calm$ the corresponding block algebra
$\gt\eq\gb\otimeS{\cal F}(\pe{\setminus}\{p_1,p_2\Ldots p_m\}) \subset \g^m$. 
The smooth sections of this bundle will be denoted by $X(\vz)$. It is 
important to realize that while the block algebras at different points $\vp$
are isomorphic as abstract \lie s, they 
are not naturally isomorphic, and the subbundle is not necessarily
a trivial bundle. Still, by construction, the \auto s $\ommuv$ restrict to 
\auto s on the subbundle of block algebras.

A crucial property of the connection \Erf6c is that it
preserves the subbundle of \bla s. Explicitly, this can be seen as follows.
The algebra \fun\ is generated algebraically by the constant
function $\fis0\df\id$ and the functions $\feS$ with $s\eq\onetom$, where
$\feS(z)\eq(z\mi z_s)_{}^{-1}$.\,%
 \futnote{When the puncture $p_m$ is at $z_m\eq\infty$, for $s\eq m$ we rather
have $\feS(z)\eq z$. The corresponding changes in the arguments below are
obvious, and we refrain from writing them down explicitly.}
By induction \wrtt `length' $\ell$ of
$f\eq\fis{i_1}\fis{i_2}\cdots\fis{i_\ell}$ and using the Leibniz rule,
it follows that this property is established as soon as it is shown to
hold for each of the functions $\feS$. Thus we consider the element 
$X\eq\xb\ot\feS$ of the \bla. For $s\eq0$ we trivially have $\dsp X\eq0$ and 
$[L_{-1}^\sssp,X]\eq0$ for all $s'\eq\onetom$. For $s\iN\{\onetom\}$ and 
$s'\nE s$
both $\dsp X$ and $[L_{-1}^\sssp,X]$ have a component only in $\g_{s'}$, namely 
  \be  [L_{-1}^\sssp,X]\ssp = \xb \oT (t\pl z_{s'}\mi z_s)^{-2}
  = - \dsp X\ssp  \Labl7y
so that again $DX\eq0$. Finally, for $s\iN\{\onetom\}$ and $s'\eq s$, the
commutator $[L_{-1}^\sssp,X]$ is as in \Erf7y, but now for $\dsp X$ the 
component in $\g_{s'}$
vanishes while there are additional contributions in all $\g_{s''}$ with
$s''\nE s$, namely $\dsp X\ssn{s''}\eq \xb \ot (t\pl z_{s''}\mi z_s)^{-2}$.
Thus in this case $D_{s'}X$ is not zero; however, it is still an element
of the block algebra \gT, namely $D_{s'}X\eq \xb \ot (z\mi z_s)_{}^{-2}$,
since we encountered precisely the local expansions of this element of \gT.

Next we consider the trivial bundle over $\calm$ with fiber given by the
tensor product $\vh\,{\equiv}\,\vhl$ \Erf vh of \ihwm s $\hil_{\Lambda_s}$ 
over the affine \lie\ \g,
  \be  \vh \Times \calm \to \calm \, . \labl6h
Again this bundle is endowed with a flat
connection $\nabla$, which acts on smooth sections $v(\vz)$ of \Erf6h as
  \be  \nabls v(\vz):= \llb \partial_s \pl L_{-1}^\ssss\lrb v(\vz)  \Labl6i
for $s\eq\onetom$; we call this connection the {\em\kzc\/}. 
And again on the bundle \Erf6h there is a fiberwise action of a certain 
discrete group. This group is the (proper) subgroup $\GF$ of \G\ that consists 
of all those elements $\muv$ of \G\ for which all associated maps $\tommuvs$ of
the tensor factors (see \erf{tommuv}) of $\vh$ are {\em endo\/}morphisms.
This group is given by
  \be  \GF = \GFL := \{ \muv\iN\G \,{\mid}\, 
  \omeuvsd(\Las)\eq\Las\;{\rm for\;all} \;s\eq\onetom\} \,.  \Labl GF
Note that by the results of subsection \ref{s.gci}, the definition of
\GFL\ depends on the \auto s $\Ommuv$ only 
through the class $[\muv]$ in \GQ\ \Erf GQ, or what is the same, through the
associated diagram \auto s $\Omeuv$.
We call \GFL\ the {\em stabilizer subgroup\/} of \G\ associated to the
weights $\Lambda_1\Ldots\Lambda_m$.
For every $\muv\iN\GFL$ we write $\Tommuv(\vz)$ for the map on 
the fiber over $\vp\iN\calm$ in the bundle \Erf6h. 

\subsection{Moduli dependence of the twisted intertwiners}

To determine how the implementation of the automorphism depends on the moduli,
we first study how the \gsauto\ $\Ommuv$ changes when the punctures are
varied (while the shift vector $\muv$ is kept fixed). 
We denote by $\vpo$ a reference point on $\calm$ and study how the
\auto\ $\Ommuv\,{\equiv}\,\Ommuv(\vz)$ differs from $\Ommuv(\vzo)$, or in other
words, what the \auto\ $\Demuv$ defined by
  \be  \Ommuv(\vz) = \Demuv(\vz;\vzo) \circ \Ommuv(\vzo)  \labl{demuv}
looks like. Note that we have to interpret $\Demuv$ as an \auto\ of the \alg\
$\gM^m$; since the block algebra, regarded
as a subalgebra of $\gM^m$, varies with the moduli, it does not
make sense to look for an automorphism of the block algebra.

For every $\so\eq\onetom$ the explicit form of the map 
$\demuv\,{\equiv}\,\demuvo$ follows directly from the formula
\Erf Au for $\ommuv\,{\equiv}\,\ommuvo$. We get $\demuv(K)\eq K$ and
  \be  \bearl
  \demuv(\hi\ot f)= \hi\ot f+ K\,\summsp\mus^i\,\Res\llb (\fiss\mi\fisso)\,f\lrb
  \,, \Nxl11
  \demuv(\ebe\ot f)= \ebe\ot f \cdot \displaystyle\prod_{s=1}^m
  (\fiss/\fisso)_{}^{-(\musp,\betab)}  \eear \Labl de
with $\fiss$ as defined in \erf{fiss}\,%
 \futnote{Here and below we suppress again the obvious modifications that
arise when the puncture $p_m$ is at $z_m\eq\infty$.}
and $\fisso(t)\df(t\pl z_\so^\kln\mi z_s^\kln)^{-1}$.

For simplicity, let us for the moment restrict our attention to the particular
case where $z_s^\kln\eq z_s$ except for some fixed value $u\iN\{\onetom\}$.
Then $\Demuv$ acts on $\gM^m$ as $\Demuv(K)\eq K$ and
  \be  \bearl  \Demuv(\hi\ot f)= \left\{ \bearll
  \hi\ot f + K\,\muu^i\,\Res\llb (\fisu\mi\fisuo)\,f\lrb &{\rm for}\ \so\nE u\,,
  \\[.3em]
  \hi\ot f + K\,\summu\mus^i\,\Res\llb (\fius\mi\fiuso)\,f\lrb
  &{\rm for}\ \so\eq u\,, \eear\right. \nxl9
  \Demuv(\ebe\ot f)= \left\{ \bearll
  \ebe\ot f \cdot (\fisu/\fisuo)_{}^{-(\muu,\betab)} &{\rm for}\ \so\nE u\,,
  \\[.3em]
  \ebe\ot f \cdot \dsty\produ (\fius/\fiuso)_{}^{-(\mus,\betab)} 
  &{\rm for}\ \so\eq u\,. \eear\right.  \eear \Labl du

Next we observe that a variation of the moduli does not change the class of the 
automorphism $\ommuv$ modulo inner automorphisms. As a consequence, $\Demuv$ is 
in fact an inner \auto\ of $\gM^m$. As can be verified by direct calculation 
(see appendix \ref{AB}), we have in fact
  \be  \Demuv = \left\{ \bearll
  \expad{(\muu,H)\ot\gsu} &{\rm for}\ \so\nE u\,, \\[.3em]
  \expad{\sum_{s\ne u}(\mus,H)\ot\hs} &{\rm for}\ \so\eq u\,,
  \eear\right.  \Labl ad
where
  \be  \gsu(t):= \ln \frac{t+z_{s_0}-z_u}{t+z_{s_0}-z^\kln_u}
  \qquad{\rm and}\qquad
  \hs(t) := \ln \frac{t+z_u - z_s}{t+z_u^\kln -z_s} \,.  \labl{gsO,hs}
(Here we choose some definite branch of the logarithm. It is readily checked
that all the results below do not depend on this choice. For $z_u\nE z^\kln_u$,
$\gsu$ can be considered as an element of $\complex\twobrac t$,
for which the constant part is defined only modulo $2\pi\ii\zet$.)

Now recall from section \ref{s.cb} that associated to the \auto\ $\ommuv$ of
$\gM^m$ there comes the map $\tommuv$ that for each tensor factor of $\vh$ is 
defined as in \erf{tommuv}.
This map is present for any $\vz\iN\calm$; moreover, according to the relation
\erf{demuv} the maps $\tommuv(\vz)$ and $\tommuv(\vzo)$ are related by an
analogous map that is associated to the inner \auto\ $\Demuv$. Using the results
\Erf t' and \Erf ad we learn that explicitly we have
  \be  \tommuvo(\vz) = \exp\llb(\muu,H)\ot\gsu+ K \fsO \lrb \circ \tommuvo(\vzo)
  \quad\ {\rm for}\;\ \so\nE u \,,  \Labl2t
\resp
  \be  \tommuvo(\vz) = \exp\llb\summu (\mus,H)\ot\hs+K\fsu\lrb \circ
  \tommuvo(\vzo) \quad\ {\rm for}\;\ \so\eq u \,.  \Labl3t
At this stage the functions $\fsO$, which are introduced here to 
take care of the fact that the implementation is only determined up to a phase, 
can still be chosen arbitrarily; we will make use of this freedom later on.

Combining these results with the twisted intertwining relations \Erf TX,
we can show that the commutator $[\nablu,\Tommuv(\vz)]$ has the local
expansions
  \be  [\nablu,\tommuvo(\vz)] \zuo
  = \Llb {-}(\muu,H)\oT\fisuo + K\, \dfsO \Lrb\, \tommuvo(\vzo) \,,  \Labl8a
where
  \be  \dfsO = 
  \Frac\partial{\partial z_u}\,\fsu\zuo - \summu \Frac{(\mus,\muu)}
  {z_u^\kln\mi z_s^\kln} \qquad{\rm for}\;\ \so\eq u\,,  \Labl5s
while $\dfsO\eq\frac\partial{\partial z_u}\,\fsO\zuo$ for $\so\nE u$.
(The derivation of \Erf8a is explained in some detail in appendix \ref{AB}.)
Collecting these local expansions, we learn that
up to central terms we simply have
  \be  [\nablu,\tommuv(\vzo)] = \Yu\, \tommuv(\vzo) \,,  \Labl0t
where
  \be  \Yu := -\,(\muu,H)\oT (z\mi z_u^\kln)^{-1}  \Labl Yu
is an element of the block algebra.
Concerning the central terms, we claim that we can employ the freedom that is
present in the choice of the functions $\fsO$ so as to achieve 
  \be  \dfsO \equiv 0 \quad\ \mbox{for all }\; \vp\in\calm \,.  \ee
Clearly, this is possible at the point $\vp\eq\vpo$, by just choosing
 \futnote{Here the freedom in the choice of the branch of the logarithm is
not completely irrelevant. But since the level is always integral while the
inner product $(\mus,\muu)$ is rational, the twisted intertwiner $\tommuvo$
is determined only up to multiplication by a root of unity, and the presence 
of this root of unity does not destroy the important property of $\tommuvo$
to have finite order.
Moreover, the precise value of the order really matters only in the case of 
fixed points, and closer inspection shows that in that case the number $(\mus,
\muu)$ is in fact always an integer so that the order is not changed at all.}
  \be  \fsO\equiv 0 \quad\ \mbox{for }\; \so\nE u \qquad{\rm and}\qquad
  \fsu =   \summu (\mus,\muu)\, \ln \frac{z_u\mi z_s}{z^\kln_u\mi z_s^\kln}
  \,.  \Labl5v

To discuss how the situation looks like globally, we first have to generalize
the formula \Erf8a to the case where all punctures may vary. By analogous
considerations as above one can check that the
generalization of \Erf2t and \Erf3t reads
  \be  \tommuvo(\vz) = \exp \llb\summs (\mus,H)\oT\gsos+K\FsO\lrb \circ
  \tommuvo(\vzo)  \Labl3s
for all $\so$, where
  \be  \gsos(t) := \ln \frac{t+z_{s_0}-z_s}{t+z_{s_0}^\kln-z^\kln_s} \,,
  \labl{gsos}
which for all $\so$ and all $s$ is a sensible Laurent series.
(Also note that the notation $\gsou$ is in agreement with the definition
\erf{gsO,hs}, and $\gus\eq\hs$;
the functions $\FsO$ are still to be determined.) The same arguments as in
appendix \ref{AB} then lead to a formula analogous to \Erf8a,
  \be  [\nablu,\tommuvo(\vz)] \zo
  = \Llb {-}(\muu,H)\oT\fisuo + K\, \dfsO \Lrb\, \tommuvo(\vzo) \,,  \Labl8d
with
  \be  \dfsO = \Frac\partial{\partial z_u}\,\FsO\zo - \summu \Frac{(\mus,\muu)}
  {z_u^\kln\mi z_s^\kln} \Labl5t
for all $\so$. Our aim is now to choose the functions $\FsO$ is such a way that 
in the tensor product $\Tommuv$ the terms proportional to $K$ in the
exponent cancel so that we are again left with an element of the block algebra.
Thus we need again $\sum_\so\dfsO \equiv 0$, i.e.
  \be  \summso \Frac\partial{\partial z_u}\,\FsO = \summu \Frac{(\mus,\muu)}
  {z_u\mi z_s}  \Labl5T
(note that the appearance of the summation is in agreement with the fact that
$\gM^m$ is obtained from a direct sum of affine \lie s by identifying the
centers). The choice
  \be  \FsO(\vz) = \half \sumss (\mub_\so,\mub_s)\, \ln
  \frac{z_\so\mi z_s}{z^\kln_\so\mi z_s^\kln}  \Labl5u
for all $\so\eq\onetom$
indeed satisfies this requirement. Note that the requirement \Erf5T determines
$\FsO$ only up to a $\vz$-independent constant; the choice made here will be 
convenient later on.\,%
 \futnote{Also note that in the general case considered here there is of
course less freedom in the choice of these functions than we had for the
analogous functions $\fsO$ in the special case treated before. In particular,
the simple choice made in \Erf5v is no longer available.}
We conclude that we can choose the phase of $\Tommuv(\vz)$ in such a manner
that the \kzc\ is preserved at every point of the moduli space.

\subsection{Transformation properties of flat sections}
  
We are now in a position to investigate the bundle of \cb s and
the transformation properties of those sections in the bundle which are flat
\wrtt \kzc. To this end we consider the trivial bundle
  \be  \vhd \Times \calm \to \calm \,,  \Labl27
where the fiber $\vhd$ is the algebraic dual of $\vh$, and in this bundle the
{\em subbundle \cbb\ of \cb s\/} which are the singlets under the (dual) action
of the block algebra. This implies that the smooth 
sections $\bl(\vz)$ of \cbb\ obey
  \be  \pair{\bl(\vz)}{X(\vz)v(\vz)} = 0  \Labl Bl
for all sections $X(\vz)$ of the bundle of block \alg s and all sections
$v(\vz)$ of the trivial bundle \Erf6h. This bundle \cbb\ is of finite rank
\cite{Ueno}. We endow the trivial bundle \Erf27 with the 
connection that is dual to the connection \erf{6i}; it restricts to a 
connection on the subbundle \cbb. We will use the term \kzc\ also for both 
the connection on the dual bundle and for its restriction to $\cbb$.
(Note, however, that a frequent convention in the literature is to reserve 
the term \kzc\ only for the connection on \cbb.)

An action of the group \GF\ on \cbb\ can be defined by
  \be  \pair{\Tommuvd(\bl(\vz))}{v(\vz)} := \pair{\bl(\vz)}{\TommuvM v(\vz)}
  \ee
for all sections $v(\vz)$ (recall that $\Tommuvd(\bl)$ is again a \cb).
We have 
  \be \bearll  \pair{\nabls(\Tommuvd\bl(\vz))}{v(\vz)} \!\!
  &= \partial_s\pair{\Tommuvd\bl(\vz)}{v(\vz)}
   - \pair{\Tommuvd\bl(\vz)}{\nabls v(\vz)} \nxl7
  &= \partial_s\pair{\bl(\vz)}{\TommuvM v(\vz)}
   - \pair{\bl(\vz)}{\TommuvM \nabls v(\vz)} \,.  \eear  \ee
Using the result \Erf0t, this can also be written as 
  \be  \hsp{-.9}  \bearll \pair{\nabls(\Tommuvd\bl(\vz))}{v(\vz)} \!\!\!
  &= \partial_s\pair{\bl(\vz)}{\TommuvM v(\vz)}
  - \pair{\bl(\vz)}{\nabls\TommuvM v(\vz)} 
     + \pair{\bl(\vz)}{\Ys\,\TommuvM v(\vz)} \nxl6
  &= \partial_s\pair{\bl(\vz)}{\TommuvM v(\vz)}
  - \pair{\bl(\vz)}{\nabls\TommuvM v(\vz)} \,,  \eear \Labl0u
where in the second line we used the fact that according to the definition
\Erf Yu, $\Ys$ is an element of the block algebra.
Note that $\Tommuv(\vz)$ depends smoothly on $\vz$, so that
$\TommuvM(\vz)v(\vz)$ is a smooth section as well.
Let now $\bl(\vz)$ be a {\em flat\/} section in \cbb,\,%
 \futnote{Sometimes in the literature the term `\cb' is reserved for such flat 
sections.} 
i.e.\ $\nabla\bl\eq0$, or more explicitly,
  \be  0 = \pair{\nabls\bl(\vz)}{v(\vz)}
  = \partial_s\pair{\bl(\vz)}{v(\vz)} - \pair{\bl(\vz)}{\nabls v(\vz)}  \ee
for all smooth sections $v(\vz)$ and all $s\eq\onetom$.
Applying this formula to the smooth section $\TommuvM(\vz)v(\vz)$, we see from
\Erf0u that
  \be  \pair{\nabls(\Tommuvd(\vz)\bl(\vz))}{v(\vz)} = 0 \,.  \ee
This means that together with $\bl(\vz)$ also the section
$\Tommuvd(\vz)\bl(\vz)$ is flat.

\subsection{Verification of the projective action of \G}

In subsection \ref{s.cb1} we have seen that for every given value $\vz$
of the moduli the maps $\Tommuvz$ that implement the automorphisms $\Ommuv$
on the $\g^m$-modules can be chosen such that they respect the group law of 
\G\ up to a two-cocycle $\epsilon$. 
Suppose we have made this choice for some point $\vzo\iN\calm$.
To extend $\Tommuv$ to other values of the moduli, we have imposed the 
requirement that the extension should be compatible with the \kzc. This 
requirement can only be satisfied when the implementation $\Tommuvz$ is chosen 
in a suitable manner.  We will now show that the specific implementation 
$\Tommuvz$ that we have already chosen above for each $\muv\iN\G$
still respects the group law of \G\ at {\em every\/} point in $\calm$
up to the {\em same\/} cocycle $\epsilon$.

To start, we recall the formula \Erf3s, i.e.\
$\tommuvo(\vz)\eq\Inner\muv(\vz;\vzo)\,\tommuvo(\vzo)$, with
  \be  \Inner\muv(\vz;\vzo) = \exp \llb\summs (\mus,H)\ot\gsos+K\FsO\lrb \,; 
  \labl{Inner}
here $\gsos(t;\vz)$ and $\FsO(\vz;\muv)$ are defined by \erf{gsos} and
\Erf5u, \resp. It follows that
  \be \bearll  \epsilon(\muv_,\nuv)\, \tomnuvo(\vz) \!\!\!
  &= \Inner{\muv+\nuv}(\vz;\vzo) \circ \tomnuvo(\vzo) \nxl8
  &= \Inner\muv(\vz;\vzo) \circ \exp\llb \summsp(\nus,H)\gsos +
     K\, [\FsO(\vz;\muv{+}\nuv) - \FsO(\vz;\muv)] \lrb \Nxl12
  & \hsp{20} \circ \tommuvo(\vzo)\,\tonnuvo(\vzo) \,.  \eear \ee
By commuting the exponential through $\tommuvo(\vzo)$ we arrive at a similar
exponential, but with $(\ommuV^\kln)^{-1}$ applied to the argument.
With the help of the identity \erf{35} we then get explicitly
  \be  \epsilon(\muv_,\nuv)\, \tomnuvo(\vz) = \Inner\muv(\vz;\vzo) \circ
  \tommuvo(\vzo) \circ \exp(\tilde Y_\so) \circ \tonnuvo(\vzo)  \ee
with 
  \be \bearll  \tilde Y_\so \!\!
  &:= \dsty \sum_s (\nus,H)\,\gsos + K\, \Llb -  \sum_{s,s'}
  (\musP,\nus)\, \Res(\fissp\gsos) \Nxl11
  & \hsp{7.5}  +\Frac12\, \dsty \sum_{s\atop s\ne\so}
    \ln\Frac{z_\so-z_s}{z_\so^\kln-z_s^\kln}
    \lLb (\muso{+}\nub_\so,\mub_s{+}\nub_s) \mi (\muso,\mub_s)
    \lRb \Lrb \nxl8
  &= \dsty \sum_s (\nus,H)\,\gsos +\Frac12\, K\, \sum_{s\atop s\ne\so}
    \ln\Frac{z_\so-z_s}{z_\so^\kln-z_s^\kln}\, (\nub_\so,\nub_s) \nxl9
  & \hsp{7.5}  +\Frac12\, K\, \dsty\sum_{s\atop s\ne\so}
    \ln\Frac{z_\so-z_s}{z_\so^\kln-z_s^\kln}\, \lLb (\nub_\so,\mub_s)
    - (\muso,\nub_s) \lRb \,.  \eear \ee
Upon exponentiation, the terms in the first line of the last expression
just yield the correct factor
$\Inner\nuv(\vz;\vzo)$, while the rest of the terms amount to an
additional phase. Now this is the situation at the puncture $\so$; 
taking into account that the centers of the affine \alg s \gS\ are
identified in $\g^m$, we finally have to add
up the prefactors of $K$ for all insertion points $\so\eq\onetom$. Doing so,
the two sums in the prefactor cancel each other,
  \be  \sum_{\so,s\atop s\ne\so} \ln\Frac{z_\so-z_s}{z_\so^\kln-z_s^\kln}\,
  \lLb (\muso,\nub_s) - (\nub_\so,\mub_s) \lRb = 0 \,.  \ee
As a consequence, we have
  \be \bearll  \epsilon(\muv_,\nuv)\, \Tomnuv(\vz)  \!\!
  &= \INner{\muv+\nuv}(\vz;\vzo) \, \Tomnuv(\vzo) \nxl6
  &= \INner\muv(\vz;\vzo)\Tommuv(\vzo) \circ \INner\nuv(\vz;\vzo) \Tonnuv(\vzo) 
  = \Tommuv(\vz) \circ \Tonnuv(\vz) \,.  \eear \ee
We conclude that, as claimed, with our chosen implementation the group law 
of \G\ is respected at every point of the moduli space $\calm$
up to a $\vz$-independent cocycle. (Note that it is the representative cocycle
$\epsilon$ itself, and not just its cohomology class, that is independent of
the moduli.)

Moreover, using the relation
  \be \hsp{-1.1}\bearl
  \exp\llb \Frac{\ii\pi}2\,(\eals\ot\fiss +\emals\ot\fiss^{-1}) \lrb \nxl5
  \hsp{2.8} = \exp\llb{-}\Frac12\, \hals\ot\gsos\lrb \,
  \exp\llb \Frac{\ii\pi}2\,(\eals\ot\fisso+\emals\ot(\fisso)^{-1}) \lrb   \,
  \exp\llb \Frac12 \, \hals\ot\gsos\lrb \,,  \eear  \Labl3y
one can show that the preferred implementation of inner \gsauto s 
introduced in subsection \ref{s.cb1} (see \erf{primp}) obeys
  \be  \ttommuvo(\vz) = \Inner\muv(\vz;\vzo)\,\ttommuvo(\vzo)  \Labl3z
precisely as in \Erf3s, which implies
that at every point in the moduli space this implementation
is compatible with the \kzc. (For details, we refer to appendix \ref{AD}.)
We conclude in particular that in each fiber of the bundle \cbb\ the map 
$\Tommuvd$ has finite order. 

\sect{Fixed point resolution}\label{s.tf}

\subsection{Fixed points}

In the previous section we have constructed a projective action of the group
\G\ on the tensor product $\vh$ in such a way that for each $\muv\iN\GO$ 
the twisted intertwiner $\Tommuv$ is represented by the product of 
exponentials of elements in the block algebra $\gt\,{\equiv}\,\gt(\vzo)$.
As seen in subsection \ref{s.gci} this has in particular the consequence
that in all fibers over the moduli space $\calm_m$ the induced maps on the 
space $\cbs\,{\equiv}\,\cbs(\vz)$ of \cb s have finite order and
realize, modulo the fixed cocycle $\epsilon$, the group law of the finite 
abelian group $\G/\GO$. 
   
A particularly interesting situation arises when the automorphism $\ommuv$ is 
not inner, but still does not change the isomorphism class of a module \hl. More
precisely, given an $m$-tuple $\vec\Lambda$ of integrable weights of the 
affine \lie\ \g, we associate to it the subgroup of \G\ 
that leaves each of the \ihwm s $\hil_{\Lambda_s}$ invariant up to
isomorphism. This subgroup is precisely the stabilizer \GFL\ of $\vec\Lambda$
as defined in \Erf GF. For every $m$-tuple $\vec\Lambda$ the stabilizer \GFL\
definitely contains \GO\ as a subgroup. If it is larger than \GO, then it also 
contains elements which do not necessarily act as a multiple of the identity 
on the blocks. In this case we call the $m$-tuple
$\vec\Lambda$ of \g-weights a {\em fixed point\/}.

Now the cocycle $\epsilon\,{\equiv}\,\epsilon_{\vec\Lambda}$ on 
$\GG/\GGO$ induces a cocycle on the subgroup $\GFL/\GO$, 
which we denote by the same symbol. Further, when applied to fixed points, the 
results of subsection \ref{s.gci} tell us that each fiber of the vector bundle
$\cbb$ of \cb s can be split into finitely many 
subspaces $\cbs^\psi$ that are invariant under the projective action of the 
group \GFL, or rather, of the quotient by its subgroup \GO. These 
invariant subspaces, whose dimensions may be larger than one,
are in correspondence with the irreducible \rep s of the 
twisted group \alg\ \CEL\ of the finite group $\GFL/\GO$. Thus\,%
 \futnote{See e.g.\ \cite{KArp4} for the \rep\ theory of twisted group \alg s.}
the label $\psi$ of the \eigenspace\ $\cbs^\psi$ can be taken to be
a character of the center
\CELC\ of \CEL, which in turn (compare the remarks around formula \erf{REG}) 
is the group \alg\ $\CELC\eq\complex(\REGL)$ of the subgroup $\REGL$ of regular
elements of $\GFL/\GGO$; thus in short, $\psi\iN(\REGL)^\dual_{}$.
Now the \kzc\ is preserved under the 
map $\Tommuvd$; therefore it restricts to the various \eigenspace s and they fit
together into sub-vector bundles of $\cbb$.
In particular, the dimensions of the \eigenspace s $\cbs^\psi$ do not depend on 
the moduli. Hence as soon as $\REGL$ is non-trivial,
the bundle \cbb\ of \cb s is reducible (as a 
vector bundle). To actually establish this fact, we have to show that the 
ranks of at least two subbundles $\cbb^\psi$ are non-zero, which in fact 
follows from the conjectured formula for the rank to be discussed below.
We refer to the decomposition of the bundle \cbb\ into the subbundles 
of \eigenspace s under $\Tommuv$ as {\em fixed point resolution\/}.
Our results imply that the \kzc\ consistently restricts to these 
subbundles; thus even after fixed point resolution we are still given a 
\kzc. 
The present situation should be compared with the situation in coset 
\cfts, where it is known \cite{fusS4} that the characters, i.e.\ the zero-point
blocks on the torus, of fixed points decompose in a similar way
under the action of an outer automorphism. This structural analogy
should also explain the striking similarities in the modular matrices
for extension modular invariants \cite{fusS6} and coset \cfts\ \cite{fusS4}.
 
We pause for a side remark. One might be tempted to speculate that the
converse is also true, i.e.\ that the bundle \cbb\ splits into
a direct sum of subbundles 
only if fixed points are involved.
This, however, does not seem to be true, as the following example at
higher genus shows. The bundle of zero-point blocks on the torus is
given by the characters of the theory, and the representation of the mapping
class group on this bundle is just given by the usual modular group
representation on the
characters. This bundle definitely does not involve fixed points. On the other
hand it is known that the representation of the modular group is
reducible if the theory contains non-trivial simple currents. (One well
known example for this phenomenon is the fact that in superconformal field 
theories the character-valued indices of fields in the Ramond sector span
a closed subspace under modular transformations.)
 
\subsection{Trace formul\ae}\label{s.Tf}

We have seen that the dimensions of the \eigenspace s $\cbs^\psi$ 
do not depend on the moduli. In this subsection we present a conjecture
for a general formula for these dimensions. Via Fourier transformation over
the finite abelian group $\REGL$ of regular elements of $\GFL/\GO$,
these dimensions are related to the traces 
of the implementing maps $\Tommuvd$ on the fibers. More precisely, we need to
choose a representative $\muv\iN\GFL$ for every element $\vec\omega$ of $\REGL$;
the result does not depend on the choice of representative modulo \GO, because 
the map $\Tommuvd$ depends on $\muv\iN\GFL$ only through its class $\vec\omega$.
Now recall that $\omeuvs$ denotes the diagram \auto\ that is in the same class
as $\ommuvs$. We have also seen that the latter is in the 
same class as the ordinary single-shift \auto\ $\ommuos$; accordingly, 
instead of $\omeuvs$ we may also use the notation $\omeus$ or, for brevity,
$\Oms$. In this notation, the condition that $\sum_{s=1}^m\mus\eq0$ tells us 
that we have
  \be  \prod_{s=1}^m \Oms = \id \,.  \Labl 73
Moreover, as diagram automorphisms are in one-to-one correspondence with
classes of outer automorphisms, we identify the $m$-tuple $\vec\omega$
of outer automorphisms with the corresponding element of the group $\REGL$.

Also recall that the implementing maps $\Tommuvd$ were defined only up to a 
phase. The invariant contents of the conjecture we are going to spell out
consists of a formula for the 
dimensions of the invariant subspaces $\cbs^\psi$. However,
it will be convenient not to write down directly this formula for
the dimension of $\cbs^\psi$, but to present instead the trace 
  \be  \Tr_{\vec\Lambda;\vec\omega} \equiv
  \Tr_{\vec\Lambda;\vec\sigma^{}_{\vec{\Bar\mu}}} :=
  {\rm tr}_{{\cal B}_{\Vec\Lambda}} \, \Tommuvd  \Labl Tr
for a given choice of implementation $\Tommuvd$. Of course, unlike the
dimensions of the invariant subspaces themselves, these numbers do depend on
the chosen implementation. 
For a definite choice of this phase, the relation between the
dimensions $\dim \cbs^\psi$ and traces reads
  \be  \dim \cbs^\psi = |\REGL|_{}^{-1}\, \sum_{\vec\omega\in\REGL}
  \psi^*(\vec\omega)\, \Tr_{\vec\Lambda;\vec\omega}\,.  \Labl9t
      The facts that the \rhs\ has an independent meaning and that
      the traces on the \lhs\ depend on the implementation
are reconciled by the following
observation. First, we do require that the maps $\Tommuvd$ realize the group 
structure of \G\ projectively and that they are the identity for all 
$\muv\iN\GGO$. This already restricts the choice of possible scalar factors,
but still leaves some indeterminacy. Indeed, the remaining freedom consists of
modifying the implementing maps by phases that furnish a character $\varphi$
of the group $\REGL$. 
In short, once we fix the choice of implementation in such a way that our
requirements are satisfied, we can label the eigenspaces by characters $\psi$
of $\REGL$, but when making an allowed change in the implementation, the
labelling of the eigenspaces is by different characters $\psi'\eq\psi\varphi$
of $\REGL$.

Having explained the dependence of the traces \Erf Tr on the choice of 
implementation, we are now in a position to present our conjecture for these
numbers. The results of \cite{fusS6} for the $S$-matrix of integer spin simple
current modular invariants suggest, when combined with the ordinary Verlinde 
formula, that there exists an allowed implementation for which the traces in 
the case $m\eq3$ are given by
  \be  \Tr_{\vec\Lambda;\vec\omega} = 
  \sum_\Lap \frac{ S^{\omega_1}_{\La1,\Lap} S^{\omega_2}_{\La2,\Lap}
  S^{\omega_3}_{\La3,\Lap}} {S_{\vac,\Lap}} \,.  \labl{conj1}
Here the summation extends over all integrable \g-weights at level $\kv$ which
are fixed under each of the permutations $\omd_s$ for $s\eq1,2,3$;
$\vac$ is the label for the vacuum primary field, while
$S_{\Lambda,\Lambda'}$ denotes the entries of the modular $S$-matrix
of the \wzwt\ based on \g, which is given by the \kpf\ \cite{kape3}
  \be  S_{\Lambda,\Lambda'}^{} = {\cal N} \sum_{\bar w\in\overline W}
  {\rm sign}(\bar w) \, \exp\lLb -\Frac{2\pi\ii}{\kV+\gV}\,
  (\bar w(\bar\Lambda+\bar\rho),\bar\Lambda'+\bar\rho) \lRb\,. \labl8k
And finally, $S^{\omega}_{\La,\Lap}$ are the entries of the modular $S$-matrix 
for some other \wzwt. Namely, to each pair consisting of an affine \lie\ \g\
and a diagram automorphism $\omega$ of \g\ one can associate another
affine \lie\ $\g^\omega$, the so-called orbit \lie;
$S^{\omega}_{}$ is just given by the \kpf\ for $\g^\omega$. (For more details, 
in particular on how the fixed point weights $\Lambda_s$ are to be interpreted 
as weights of $\g^\omega_s$, see \cite{fusS3,furs}. For convenience, 
we have also listed in table 1 the orbit \lie s\,%
\futnote{$\Bntilde$ stands for the unique series of twisted affine \lie s whose 
characters furnish a module of the modular group and which have simple roots 
of three different lengths.}
of untwisted affine \lie s \wrt all relevant diagram \auto s.) 
\begin{table}[bhtp]\caption{{\em Orbit \lie s of untwisted affine \lie s}
 [22].}
\begin{center}
  \begin{tabular}{|l|c|c|l|}
  \hline &&&\\[-.9em]
  \multicolumn{1}{|c|} {\g} & \multicolumn{1}{c|}  {$\omega$} &
  \multicolumn{1}{c|}  {$N$} & \multicolumn{1}{c|} {$\gM^\omega$}
  \\[1.2mm] \hline\hline &&&\\[-2.8mm]
   $A\untw_{n-1}$  & $\omO$ & $n$ & $ \none $               \\[1.9mm]
   $A\untw_{n-1}$  & $(\omo)^{n/N} $& $N<n$ & $A\untw_{(n/N)-1}$ \\[1.9mm]
   $B\untw_{n}$    & \sicu  & 2  & $ \tilde B_{n-1}\twtw $  \\[1.9mm]
   $C\untw_2$      & \sicu  & 2  & $ A\twtw_1 $             \\[1.9mm]
   $C\untw_{2n}$   & \sicu  & 2  & $ \Bntilde $             \\[1.9mm]
   $C\untw_{2n+1}$ & \sicu  & 2  & $ C\untw_n $             \\[1.9mm]
   $D\untw_n$      & \jv    & 2  & $ C\untw_{n-2}$          \\[1.9mm]
   $D\untw_{2n}$   & \js    & 2  & $ B\untw_n $             \\[1.9mm]
   $D\untw_{2n+1}$ & \js    & 4  & $ C\untw_{n-1}$          \\[1.9mm]
   $E\untw_6$      & \sicu  & 3  & $ G\untw_2 $             \\[1.9mm]
   $E\untw_7$      & \sicu  & 2  & $ F\untw_4 $            
   \\[.4em] \hline \end{tabular} 
\end{center} \end{table}

It is worth mentioning that in a first step, formula \Erf9t should be regarded 
as an identity that holds in each fiber separately. But since the dimension
$\dim \cbs^\psi$ is independent of the choice of the fiber, by inverse Fourier 
transformation one concludes that the traces also do not depend on the point 
in moduli space over which the trace is taken.
Accordingly, there is no moduli dependence in our conjecture \erf{conj1}.

It is a remarkable empirical observation that, in all cases that have been 
checked numerically, the expression \erf{conj1} gives an integral 
(though not necessarily non-negative) result.\,%
 \futnote{The relevant calculations have been performed with the program
{\tt kac} which has been written by A.N.\ Schellekens and is available at
{\tt http:/$\!$/norma.nikhef.nl/\raisebox{-.2em}{$\tilde{\phantom.}\,
$}t58/kac.html}.\\ Helpful discussions with Bert Schellekens are gratefully
acknowledged.}
Notice that this property is much stronger than the obvious requirement that 
the dimensions $\dim \cbs^\psi$ of the \eigenspace s must be integral. 
(Also, when the constraint \Erf73 is not satisfied, then the expression 
on the \rhs\ of \erf{conj1} is zero.)

We conjecture that the formula \erf{conj1} indeed holds true and, moreover,
that it directly generalizes to more complicated cases. To prepare the ground 
for this generalization, we write \erf{conj1} in the equivalent form
  \be  \Tr_{\vec\Lambda;\vec\omega} = \sum_\Lap
  \frac{S^{\omega_1}_{\La1,\Lap}}{S_{\vac,\Lap}} \Cdot
  \frac{S^{\omega_2}_{\La2,\Lap}}{S_{\vac,\Lap}} \Cdot
  \frac{S^{\omega_3}_{\La3,\Lap}}{S_{\vac,\Lap}} \Cdot |S_{\vac,\Lap}|^2 \,. 
  \Labl3S
The generalization to all $m\gE3$ then simply consists in replacing the product
$\prod_{s=1}^3 S^{\omega_s}_{\La s,\Lap}/S_{\vac,\Lap}$ by the analogous 
product $\prod_{s=1}^m S^{\omega_s}_{\La s,\Lap}/S_{\vac,\Lap}$.

The last factor on the \rhs\ of the formula
\Erf3S should find its heuristic interpretation
as a Weyl integration volume at genus zero (compare e.g.\ the derivation of
the Verlinde formula in the framework of Chern\hy Simons theory \cite{blth}).
Accordingly, for a surface of genus $g$ we can speculate
that the exponent gets replaced by the Euler number $\chi\eq2\mi2g$,
leading to an expression of the form
  \be  \sum_\Lap |S_{\vac,\Lap}|^{2-2g} \, \prod_{s=1}^m
  \frac{S^{\omega_s}_{\La s,\Lap}}{S_{\vac,\Lap}} \,,  \labl{conj2}
which is consistent with factorization.
  
Unfortunately, a complete proof of \erf{conj1}
is not known at present and remains a challenge for future work.
Note that such a proof would in particular include a proof of the ordinary
Verlinde formula for \wzwts, namely for the special case when $\omega_s\eq\id$
for all $s$.

    \hepv{
\subsection{Curves of higher genus}

The last factor on the \rhs\ of the formula
\Erf3S should find its heuristic interpretation
as a Weyl integration volume at genus zero (compare e.g.\ the derivation of
the Verlinde formula in the framework of Chern\hy Simons theory \cite{blth}).
Accordingly, for a surface of genus $g$ we can speculate
that the exponent gets replaced by the Euler number $\chi\eq2\mi2g$. 

We have already seen in
subsection \ref{ss.hg} that with the appropriate modifications several of
our constructions should generalize to curves of arbitrary genus. In contrast,
it is not clear whether the results of section \ref{s.II} can be generalized 
as well. We will see, however, that this issue is not really essential
for our purposes. The only ambiguity 
in the implementation of an automorphism $\Ommuv$ is a phase $\psi(\muv)$. 
Every element of \GO\ is implemented on the space of \cb s by
the multiple $\psi(\muv){\cdot}\id$ of the identity, and the phase $\psi(\muv)$
must furnish a character of the free abelian group \GO. Moreover, it follows
from the prescription \erf{3s} that the character depends smoothly on the
modulus $\vz$, $\Tommuvd(\vz)\eq\psi(\muv;\vz)\,\id$.
In this context we should keep in mind that from the point of view of
\cft\ it is more natural to define the \cb s as a projective space rather 
than as a vector space, i.e.\ a global phase does not possess any physical 
meaning. Indeed, at higher genus the \kzc\ is not flat any longer, but only 
projectively flat. Accordingly all our considerations perfectly make sense
for {\em projective\/} spaces of \cb s, and hence in particular we get
a natural identification of the projective bundles obtained from the maps
$\Tommuvd$.

The eigenvalues are labelled by a moduli dependent character $\psi(\vz)$
of $\GFL$, but this character does not factorize necessarily any longer to
a character of $\REGL$. But still, these characters
must depend smoothly on the moduli, and also the corresponding \eigenspace s
depend smoothly on the moduli. In particular, the dimension of the \eigenspace s
$\cbs^\psi$ does not depend on the moduli, although the eigenvalue itself
potentially does. Still, we prefer to work with the Fourier transform 
  \be \Tr_{\vec\Lambda;\vec\omega}^g := \sum_{\psi\in(\REGL)^\dual_{\phantom I}}
  \psi(\omega)\, \dim \cbs^\psi     \ee
of the dimension, which by abuse of language we will call the trace 
of $\Tommuvd$ (actually, it can differ from the trace by a phase).

For a surface of any genus $g$ the expression for the traces analogous to 
\erf{conj1} is independent on the moduli, and we conjecture it to be
  \be  \Tr_{\vec\Lambda;\vec\omega}^g = \sum_\Lap
  |S_{\vac,\Lap}|^{2-2g} \, \prod_{s=1}^m 
  \frac{S^{\omega_s}_{\La s,\Lap}}{S_{\vac,\Lap}} \,.  \labl{conj2'}

As non-trivial evidence for
this formula, we show that it is consistent with factorization.
Namely, suppose a curve $C$ of genus $g$ develops an ordinary double point and 
that its normal resolution consists of two curves $C\sse,C\ssz$ of genus $g_1$ 
and $g_2$, \resp, with $g_1{+}g_2\eq g$. We assume that for $i\eq1,2$ the 
insertion points on $C\ssi$ correspond to integrable \g-weights 
$\vec\Lambda\ssi$
and that we want to implement an $m_i$-tuple of automorphisms 
$\vec\sigma_{\vec{\Bar\mu}}\ssi$ satisfying $\omdv{}_{}\ssi\vec\Lambda\ssi
\eq\vec\Lambda\ssi$.
According to \erf{73} we require that 
  $   \prod_{s=1}^{m_1} \omega_s\sse \cdot \prod_{s=m_1+1}^{m_1+m_2}
  \omega_s\ssz \eq \id  $  
on the curve $C$, which implies that
  \be \omega_\circ := \prod_{s=m_1+1}^{m_1+m_2}\vec\omega_s\ssz
   = \Llb\prod_{s=1}^{m_1} \vec\omega_s\sse \Lrb^{-1} \, . \labl{79}

Next we note that
  \be  \bearll 
  \Tr_{\!\vec\Lambda\sse,\Lap;\vec\omega\sse,\omega_\circ}^{g_1} {\cdot}\,
  \Tr_{\!\vec\Lambda\ssz,\Lap^+;\vec\omega\ssz,\omega_\circ^{-1}}^{g_2}
  \!\!\!\!\! &= \dsty \sum_{\kappa:\;\omega_\circ^\dual\kappa=\kappa}\!\!
  \Llb \sum_\Lap |S_{\vac,\Lap}|^{2-2g_1}
  \cdot \frac{S^{\omega_\circ}_{\kappa,\Lap}}{S_{\vac,\Lap}}\,
  \prod_{s=1}^{m_1} \frac{S^{\omega_s}_{\La s,\Lap}}{S_{\vac,\Lap}} \Lrb
  \Nxl11 & \hsp{3.2}
  \cdot \dsty \Llb \sum_\Lapp |S_{\vac,\Lapp}|^{2-2g_2}
  \cdot \frac{S^{\omega_\circ^{-1}}_{\kappa^+,\Lapp}}{S_{\vac,\Lapp}} \, 
  \prod_{s=m_1+1}^{m_1+m_2} \frac{S^{\omega_s}_{\La s,\Lapp}}{S_{\vac,\Lapp}}
  \Lrb \,.  \eear \ee
Using the fact \cite{fusS3}
that $(S^\omega)^{\rm T}_{}\eq S^{\omega^{-1}}$ as well as unitarity of this
matrix, the summation over $\kappa$ can be carried out.
Then the summation over, say, $\Lapp$ is immediate and leads to
  \be  \Tr_{\vec\Lambda\sse,\Lap;\vec\omega\sse,\omega_\circ}^{g_1} \,
  \Tr_{\vec\Lambda\ssz,\Lap^+;\vec\omega\ssz,\omega_\circ^{-1}}^{g_2}
  = \sum_\Lap |S_{\vac,\Lap}|^{2-2g_1-2g_2}\!
  \prod_{s=1}^{m_1+m_2} \frac{S^{\omega_s}_{\La s,\Lap}}{S_{\vac,\Lap}}
  = \Tr_{\vec\Lambda\sse,\vec\Lambda\ssz;\vec\omega\sse,\vec\omega\ssz}
  ^{g_1+g_2} \,.  \ee
Hence
  \be  \Tr_{\vec\Lambda\sse,\vec\Lambda\ssz;\vec\omega\sse,\vec\omega\ssz}
  ^{g_1+g_2} = \sum_{\Lao\;{\rm fix}} 
  \Tr_{\vec\Lambda\sse,\Lao; \vec\omega\sse,\omega_\circ}^{g_1} \cdot
  \Tr_{\vec\Lambda\ssz,\Laod;\vec\omega\ssz,\omega_\circ^{-1}}^{g_2}  \labl{pr}
holds, where the sum is over fixed points of $\omega_\circ$. Note that 
as a consequence of \erf{79} the product
of the automorphisms on each of the two components $C\ssi$ of the normalization
separately fulfils the constraint \erf{73}. An important
consequence of \erf{pr} is that all traces for any genus will be integral
if the traces \erf{conj1}
on the three-point blocks on a curve of genus zero are integral.
We also expect the factorization rule \erf{pr} to play in important role
for a rigorous proof of the consistency of the boundary conditions described
in \cite{fuSc5} on higher genus surfaces.

Apart from the factorization property, there is another
aspect of the formula \erf{conj2} which may be regarded as evidence that the
conjecture is indeed correct. Namely, the formula expresses the traces 
exclusively through matrices, namely $S\,{\equiv}\,S^\id$ and the various 
$S^\omega$, which all share the
standard properties (such as furnishing a unitary \rep\ of the modular group
and satisfying $S^\omega_{\vac,\Lambda}\,{>}\,0$) that are familiar from \cft.
It may therefore be expected 
that the \eigenspace s are the correct building blocks for the spaces of
\cb s of \wzwts\ based on non-simply connected groups, and that
as a consequence of the properties of the modular matrices $S^\omega$
the Verlinde formula will hold automatically in these \cfts\ as well. 

Unfortunately, a complete proof of 
\erf{conj2} is not known at present and remains a challenge for future work. 
Note that such a proof would in particular include a proof of the ordinary 
Verlinde formula for \wzwts, namely for the special case when $\omega_s\eq\id$ 
for all $s$.
}

\newpage
\appendix
\sect{The Virasoro \alg}\label{AV}

In this appendix we collect some information about \gsauto s of the \vira\
and its semidirect sum with the affine \lie\ \g.

We first show that for an arbitrary \auto\ $\sigma$ of an untwisted affine 
\lie\ \g, the extension to the semi-direct sum with the Virasoro
algebra is unique, if it exists.
To see this, consider two maps $\sigma_i{:}\ \vir\,{\to}\,\vir\splus\g$,
$i\eq1,2$, such that both $(\sigma,\sigma_1)$ and $(\sigma,\sigma_2)$
furnish an \auto\ of the semi-direct sum $\vir\splus\g$. Then we have
  \be  [\sigma_1(L_n),\sigma(x_m)] = -m\, \sigma(x_{m+n})
  = [\sigma_2(L_n),\sigma(x_m)] \,,  \ee
from which we learn that for every $n$ the combination $\sigma_1(L_n)\mi
\sigma_2(L_n)$ commutes with all of \g, so that
$\sigma_1(L_n)\mi\sigma_2(L_n)\eq\xi_n K\pl\eta_n C$ with $\xi_n,\eta_n\iN
\complex$ for $n\iN\zet$. But this in turn implies that  
$[\sigma_1(L_n),\sigma_1(L_m)]\eq[\sigma_2(L_n),\sigma_2(L_m)]$, or more
explicitly
  \be  (n-m)\, \sigma_1(L_{n+m}) + \Frac1{24}\, (n^3-n)\,\delta_{n+m,0}\,C
  = (n-m)\, \sigma_2(L_{n+m}) + \Frac1{24}\, (n^3-n)\,\delta_{n+m,0}\,C \,.  \ee
This finally implies that $\xi_n\eq0\eq\eta_n$ for all $n$, and hence
$\sigma_1\eq\sigma_2$ as claimed.

Next we check that the prescription \Erf Ln indeed provides us with
an extension of the \auto\ $\ommuvo$ of \g\ as defined in \Erf Au. We first 
verify that the relation \Erf lx is preserved when $\xb\eq H^i$. We have
  \be \bearll  [\ommuvo(L_n),\ommuvo(H^i\ot f)] \!\!
  &= [L_n\pl\sumlz\sum_s(\mus,H_{n-\ell})\ot\Res(t^\ell\fiss),H^i\ot f] \nxl7
  &= -H^i\oT t^{n+1}\diff f + K \dsty \sumlz \sum_s \mus^i\, \Res(t^\ell\fiss)\,
  \Res(\rmd t^{n-\ell}\, f) \,.  \eear \ee
Integrating by parts within the second residue and using the identity
  \be  \Res(fg) = \sumlz \Res(t^{\ell-1}f)\, \Res(t^{-\ell}g) \qquad{\rm for}
  \quad f,g\iN\complex\twobrac t  \Labl fg
(which follows immediately by substituting $f$ and $g$ by their Laurent 
expansions), this reduces to
  \be \bearll  [\ommuvo(L_n),\ommuvo(H^i\ot f)] \!\!
  &= -H^i\oT t^{n+1}\diff f - K \sum_s \mus^i\, \Res(t^{n+1}\fiss\,\diff f)\nxl7
  &= - \ommuvo(H^i\ot t^{n+1}\diff f) = \ommuvo\llb [L_n,H^i\ot f] \lrb \,,
  \eear \ee
which is the desired result.
Similarly, with the help of the identities \Erf1f and \Erf1g we calculate
  \be \bearll  [\ommuvo(L_n),\ommuvo(\eal\ot f)] \!\!
  &= [L_n\pl\sumlz\sum_s(\mus,H_{n-\ell})\,\Res(t^\ell\fiss),\eal\ot
     f\prd\alphab] \nxl3
  &= - \eal \oT t^{n+1}\diff(f\prd\alphab) + \dsty \sumlz \sum_s(\mus,\alphab)\,
     \eal \oT t^{n-\ell}f\prd\alphab\,\Res(t^\ell\fiss) \nxl7
  &= - \eal \oT t^{n+1}\diff(f\prd\alphab) + \eal\oT
     \,f\prd\alphab\, \dsty \sum_s(\mus,\alphab) \fiss\,t^{n+1}  \nxl9
  &= - \eal \oT t^{n+1}\diff(f\prd\alphab) + \eal\oT
     \,t^{n+1}f\,\diff\prd\alphab 
   = - \eal \oT t^{n+1}\,(\diff f)\,\prd\alphab  \nxl5
  &= - \ommuvo(\eal\ot t^{n+1}\diff f)
   = \ommuvo\llb [L_n,\eal\ot f] \lrb \,.   \eear \ee
Finally we check that we are indeed dealing with an \auto\ of $\vir_\so$.
We obtain
  \be \hsp{-1.4}\bearll  [\ommuvo\!(L_n),\ommuvo\!(L_m)] \!\!\!
  &= [L_n\pl\sumlz\sum_s(\mus,H_{n-\ell})\ot\Res(t^\ell\fiss), \nxl9&\hsp{8.8}
     L_m\pl\sumLz\sum_{s'}(\musP,H_{m-\ell'})\ot\Res(t^{\ell'}\fisP)] \nxl7 
  &= (n\mi m)\,L_{n+m} + \Frac1{24}\,(n^3\mi n)\, \delta_{n+m,0}\,C  \nxl7
  &\hsp{.8} + \dsty\sumLz\sum_{s'}(\musP,H_{m-\ell'+n})\oT\Res(t^{\ell'}\fisP)
     \,[-(m\mi\ell')] \nxl7
  &\hsp{.8} - \dsty \sumlz\sum_s(\mus,H_{n-\ell+m})\oT\Res(t^\ell\fiss)
     \,[-(n\mi\ell)] \nxl7
  &\hsp{.8} + K\! \dsty \sumLL\!\sum_{s,s'} (\mus{,}\musP)\,(n{-}\ell)
     \delta_{n-\ell+m-\ell',0}\,\Res(t^\ell\fiss)\,\Res(t^{\ell'}\!\fisP)  \,.
  \eear \ee
The terms in the second and third 
line combine to $(n\mi m) \sumlz\sum_s(\mus,H_{n+m-\ell})\ot\Res(t^\ell\fiss)$.
The central term in the fourth line can be rewritten as
  \be \bearl  K \dsty \sumlL\sum_{s,s'} (\mus,\musP)\,
  \llb \Frac{n-m}2 + \Frac{\ell'-\ell}2 \lrb\,
  \delta_{n+m-\ell-\ell',0}\,\Res(t^\ell\fiss)\,\Res(t^{\ell'}\fisP) \nxl7\hsp8
  = \Frac{n-m}2\,K \dsty \sum_{s,s'} (\mus,\musP)\, \sumlz \Res(t^\ell\fiss)\,
    \Res(t^{n+m-\ell}\fisP) \nxl7\hsp8
  = \Frac{n-m}2\,K \dsty \sum_{s,s'} (\mus,\musP)\, \sumlz \Res(t^{n+m+1}\fiss
    \fisP) \,, \eear \ee
where in the first step we used the fact that except for the explicit factor
$(\ell'\mi\ell)/2$ the expression is symmetric in $\ell$ and $\ell'$, and in 
the second step we employed again the identity \Erf1g.
Collecting all terms, we see that indeed
  \be  [\ommuvo(L_n),\ommuvo(L_m)] = \ommuvo\llb[L_n,L_m]\lrb  \ee
as required.

\sect{Inner \gsauto s}\label{AA}

Here we collect some details about the derivation of some of the results
stated in section \ref{s.II}.
Let us consider the map $\adxt$ for the element
  \be  \xt = \eal \ot \fe + \emal \ot \fz   \Labl xt
of the \bla\ \gT. We have
  \be \bearl
  \adxt (\hi\ot f) = - \alphab^i\,(\eal\ot\fe f - \emal\ot\fz f) \,,\\[.7em]
  \adxT2(\hi\ot f) = 2 \alphab^i\,(\alphav,H) \oT\fe\fz f \,,\\[.7em]
  \adxT3(\hi\ot f) = -4\alphab^i\,(\eal\ot\fe^2\fz f - \emal\ot\fe\fz^2 f)
  \eear \labl{12}
etc., and hence
  \be \bearl
  \expad{\xi\xt}(\hi\ot f) = \hi\ot f
  - \half\alphab^i\,(\eal\ot\fe - \emal\ot\fz)\,
    \sinh(2\xi\sqrt\fez)\,(\fez)^{-1/2}\,f  \\[.7em]\hsp{11.3}
  + \half\alphab^i\,(\alphav,H) \oT [\cosh(2\xi\sqrt\fez) -1] \, f \,.
  \eear \Labl13
In the special case where $\fz\eq(\fe)^{-1}$ (recall that this restricts 
$\fe\iN\fun$ to lie in the subalgebra $\funz$), this reduces to
  \be  \expad{(\ipi/2)\xt}(\hi\ot f) = (\bar w_\alphab(H))^i \oT f\,,  \Labl B4
where $\bar w_\alphab(H)\eq H\mi(\alphav,H)\alphab$ is the image of $H$
under the Weyl reflection $\bar w_\alphab$.
Note that this result does not depend on the function $\fe$ at all; as a
consequence, by making use of $\bar w_\alphab^2\eq\id$
one immediately deduces the relation \erf{omH}.

When applying $\adxt$ to $\ebe\ot f\in\gt$, we distinguish between several
cases. First assume that $\betab\eq{\pm}\alphab$. Then
  \be \bearl
  \expad{\xi\xt}(\epmal\ot f) = \epmal\ot f \mp \half\, (\alphav,H) \oT
    \sinh(2\xi\sqrt\fez)\, (\fez)^{-1/2}\,\fmp f  \\[.7em]\hsp{12.55}
  \pm \half\, (\eal\ot\fe - \emal\ot\fz)\,
    [\cosh(2\xi\sqrt\fez) -1] \, (\fpm)^{-1} f\,. \eear \Labl0a
Similarly, for $\betab\nE{\pm}\alphab$ we find the following.
When neither $\betab\pl\alphab$ nor $\betab\mi\alphab$ is a \gb-root, we
simply have $\adxt(\ebe\ot f)\eq0$. When either $\betab\pl\alphab$ or
$\betab\mi\alphab$ (but not both) is a \gb-root, while
$\betab\pm2\alphab$ is not a \gb-root, then we obtain
  \be  \bearll
  \expad{\xi\xt}(\ebe\ot f) = \!\!& \ebe\ot \cosh(\xi\sqrt{\eta\fez})\,f 
  \\[.5em] & +\, \e{\pm\alphab}\betab\,\ebepm\ot
  \sinh(\xi\sqrt{\eta\fez}) \,(\eta\fez)^{-1/2}\,\fpm f \,, \eear  \Labl0b
where $\eta\equiv\eta_{\pm\alphab}\df\e{\pm\alphab}\betab\e{\mp\alphab}{\betab
\pm\alphab}$, with $\e\alphab\betab$ structure constants of the \hsa\ \gb. 
Actually, from the fact that
the $\alphab$-string through $\betab$ has only two elements, it follows
that $\e{\pm\alphab}\betab$ and $\e{\mp\alphab}{\betab\pm\alphab}$ can only
take the values $\pm1$, and that
$\betab$ and $\betab\pm\alphab$ have the same length, so that the general
identity $\fraC{(\betab\pm\alphab,\betab\pm\alphab)}{(\betab,\betab)}
\eq\fraC{\e{\pm\alphab}\betab}{\e{\mp\alphab}{\betab\pm\alphab}}$
tells us that $\eta\eq1$. 
When the $\alphab$-string through $\betab$ has more elements,
then the calculations become still a bit more lengthy.
We refrain from describing all different possibilities, because the
calculations remain straightforward. As an illustration, let
$\betab\pl\alphab$ and $\betab\pl2\alphab$, but neither $\betab\mi\alphab$
nor $\betab\pl3\alphab$ be \gb-roots; then we have
  \be \bearl
  \adxt (\ebe\ot f) = \e\alphab\betab\,\ealbe\oT \fe f\,, \\[.7em]
  \adxT2(\ebe\ot f) = \e\alphab\betab\, (\e\alphab{\alphab+\betab}\,\ezalbe\oT
    \fe + \e{-\alphab}{\alphab+\betab}\,\ebe\oT \fz)\,\fe f\,, \\[.7em]
  \adxT3(\ebe\ot f) = \eta'\e\alphab\betab\,\ealbe\oT \fe^2\fz f  \eear \ee
with $\eta'\df\e\alphab{\alphab+\betab}\e{-\alphab}{2\alphab+\betab}\pl
\e{-\alphab}{\alphab+\betab}\e\alphab\betab$, leading to
  \be  \bearll
  \expad{\xi\xt}(\ebe\ot f) = \!\!& \ebe\ot f 
  +\e\alphab\betab\,\ealbe\oT\sinh(\xi\sqrt{\eta'\fez})\,(\eta'\fez)^{-1/2}\,
   \fe f \\[.5em] &
  +\, \e\alphab\betab\, (\e\alphab{\alphab+\betab}\,\ezalbe\oT
    \fe + \e{-\alphab}{\alphab+\betab}\,\ebe\oT \fz)\,
  \\[.4em] &  \hsp{5.5}
  [\cosh(\xi\sqrt{\eta'\fez})\mi1] \,(\eta'\fez)^{-1}\,\fe f\,. \eear  \Labl0d

In the special case $\fez\eq1$ from \Erf0a we get 
  \be  \expad{(\ipi/2)\xt}(\epmal\ot f) = \epmal\ot f
  \mp\, (\eal\ot\fe - \emal\ot\fz)\, (\fpm)^{-1} f = \empal\ot \fmp^2 f  \Labl2a
for $\betab\eq{\pm}\alphab$.
When in addition $\yt\iN\gT$ is of the same form as $\xt$, but
with $\fpm$ replaced by $\gpm\iN\funz$, then it follows from \Erf2a that
  \be  \E XY{\ipi/2}(\epmal\ot f)
  = \epmal\ot (\fe)^{\mp2}(\gP)^{\pm2}f \,.  \labl{omE2}
Analogously, the special case $\fez\eq1$ of \Erf0b yields
  \be  \expad{(\ipi/2)\xt}(\ebe\ot f)
  = \ii\, \e{\pm\alphab}\betab\,\ebepm\ot\fpm f \,.  \Labl2b
Similarly, from \Erf0d one obtains (using also the fact that
$\betab$ and $\betab\pl2\alphab$ must be long roots while $\alphab$ and 
$\betab\pl\alphab$ must be short roots, which implies the identities
  \be \bearl
  \e\alphab{\alphab+\betab}\e{-\alphab}{2\alphab+\betab}
  = (\e\alphab{\alphab+\betab})^2\cdot \Frac{(\alphab+\betab,\alphab+\betab)}
    {(2\alphab+\betab,2\alphab+\betab)} = 2^2\cdot\Frac12 = 2 \,, \nxl6 
  \e\alphab\betab\e{-\alphab}{\alphab+\betab}
  = (\e\alphab\betab)^2\cdot \Frac{(\betab,\betab)}
    {(\alphab+\betab,\alphab+\betab)} =1^2\cdot 2 =2 \,, \nxl6
  |\e\alphab\betab\e{\alphab}{\alphab+\betab}| = 2 \,,
  \eear \ee
so that in particular $\eta'\eq4$),
  \be  \expad{(\ipi/2)\xt}(\ebe\ot f) = 
  \pm\,\ezalbe\oT \fe^2 f \,. \Labl8f
Taking again also $\yt$ of the form described above, the relations
\Erf2b and \Erf8f lead to
  \be  \E XY{\ipi/2}(\ebe\ot f) = \ebe\ot \fpm \gmp f 
  \qquad{\rm and}\qquad
       \E XY{\ipi/2}(\ebe\ot f) = \ebe\ot(f_+g_-)^2 f \,, \Labl2c
\resp.

For the remaining cases, the calculations are completely parallel and lead
to results analogous to \Erf2c, with $\fpm \gmp$ raised to the power
$-(\alphav,\betab)$. This finally leads to the formula \erf{omE} of the main
text.

Next we comment on the analogous calculations for the case where the
functions $f$ are Laurent series (in the local coordinate $\z s$
\resp\ the formal variable $t$). First we compute that, for $\fez\eq1$,
the second formula in \erf{12} changes to
  \be  \adxT2(\hi\ot f) = 2 \alphab^i\,(\alphav,H) \oT\fe\fz f 
  - 2 \alphab^i K\, \killing\, \Res(f_+ f\, \rmdd f_-) \,,  \ee
so that \erf{13} acquires an additional contribution proportional to the
central element,
  \be  -\half\alphaV^i K\, \Res(\fe f \rmdd \fz)\, (\cosh(2\xi)-1)) \,,  
  \labl{B15}
which when specialized to $\xi\eq\ii\pi/2$ becomes $\alphaV^i K \,\Res(\fe f
\rmdd \fz)$. Performing the same calculation also with the analogous element 
$Y$ in place of $X$,
we then arrive at the formula \erf{omH2}.

In the calculation of $\expad{\xi\xt}(\ebe\ot f)$, a change occurs only for
$\betab\eq{\pm}\alphab$; in this case we get
  \be \bearl
  \adxt (\epmal\ot f) = \mp\,(\alphav,H) \oT\fmp f 
  - K\, \killing\, \Res (f_\mp \rmdd f)\,,\\[.7em]
  \adxT2(\epmal\ot f) = \pm2\, (\eal\ot\fe - \emal\ot\fz)\,\fmp f \,,\\[.7em]
  \adxT3(\epmal\ot f) = \mp4\, (\alphav,H) \oT \fmp f 
   - 4 K\,\killing\, \Res (f_\mp \rmdd f)\,, \eear \ee
etc.\ The residue terms add up to $-\Halfkilling K\, \Res (f_\mp \rmdd f)\, 
\sinh(2\xi)$, which vanishes for $\xi\eq\ii\pi/2$, while the other terms
reproduce \Erf0b.

Finally let us compute the action of $\adxt$ on the Virasoro generators $L_n$.
By \Erf lx we find, for $\fez\eq1$,
  \be \bearl
  \adxt (L_n) = E^\alphab \oT t^{n+1}\rmdd\fe 
                + E^{-\alphab} \oT t^{n+1}\rmdd\fz \,,\\[.5em]
  \adxT2(L_n) = - 2\,(\alphav,H^i)\oT\fz\rmdd\fe\,t^{n+1} + 2\,\killing K \,
                \Res(t^{n+1}\rmdd\fe\rmdd\fz) \,,\\[.8em]
  \adxT3(L_n) = 4\, (E^\alphab \oT t^{n+1}\rmdd\fe 
                + E^{-\alphab} \oT t^{n+1}\rmdd\fz) \,, \eear \ee
and hence
  \be  \hsp{-.8}\bearll
  \expad{\xi\xt}(L_n) \!\!\! &= L_n + \half \sinh(2\xi)\,
  (E^\alphab\ot t^{n+1}\rmdd\fe+E^{-\alphab}\ot t^{n+1}\rmdd\fz) \nxl7&\hsp{.8}
  - \half[\cosh(2\xi)\mi1]\, \llb (\alphav,H)\oT \fz\rmdd\fe\, t^{n+1} \mi
    \killing K\, \Res(t^{n+1}\rmdd\fe\rmdd\fz) \lrb  \,.  \eear \ee
Specializing to $\xi\eq\ii\pi/2$, we have
  \be  \expad{(\ipi/2)\xt}(L_n) = L_n + (\alphav,H)\oT \fz\rmdd\fe\, t^{n+1} -
    \killing K\, \Res(t^{n+1}\rmdd\fe\rmdd\fz)  \,.  \ee

\sect{On the moduli dependence of inner automorphisms}\label{AB}

Our first aim in this appendix is to check the formula \Erf ad, which asserts
that the \auto\ $\demuv$ acting as in \Erf du can be represented as
$\demuv\eq\expad{(\muu,H)\ot\gsu}$ for $\so\nE u$ and as
$\demuv\eq\expad{\sum_{s\ne u}(\mus,H)\ot\hs}$ for $\so\eq u$,
with $\gsu$ and $\hs$ as defined in \erf{gsO,hs}. First, definitely
  \be  \expad{(\muu,H)\ot\gsu}(K) = K = \expad{(\mus,H)\ot\hs}(K) \,.  \ee
Second, using the relations $\,\rmd\gsu\eq\fisu\mi\fisuo$ and
  \be[ (\muu,H)\oT\gsu,H^i\ot f] = \muu^i\, \Res(\rmd\gsO\,f)\, K  \ee
together with $[K,\cdot\,]\eq0$, we have
  \be  \expad{(\muu,H)\ot\gsu}(\hi\ot f) =
  \hi\ot f + K\,\muu^i\,\Res\llb (\fisu\mi\fisuo)\,f\lrb \,,  \ee
and similarly
  \be \expad{(\mus,H)\ot\hs}(\hi\ot f) =
  \hi\ot f + K\,\mus^i\,\Res\llb (\fius\mi\fiuso)\,f\lrb \,.  \ee
And finally, by exponentiating the identity
  $[(\muu,H)\ot\gsu,\ebe\ot f]\eq(\muu,\betab)\ebe\ot\gsu f$
we obtain
  \be  \expad{(\muu,H)\ot\gsu}(\ebe\ot f)
  = \ebe\oT \exp((\muu,\betab)\,\gsu)\, f = \ebe\oT f\cdot
  \LlB \Frac{t+z_{s_0}-z_u}{t+z_{s_0}-z^\kln_u} \LrB ^{(\muu,\betab)} \,, \ee
and analogously
  \be  \expad{(\mus,H)\ot\hs}(\ebe\ot f) = \ebe\oT f\cdot
  \LlB \Frac{t+z_u - z_s}{t+z_u^\kln -z_s} \LrB ^{(\mus,\betab)} \,. \ee
Putting these results together we see that $\demuv$ is indeed given by \Erf ad.

Next we derive the formula \Erf8a for the commutator
$[\nablu,\tommuvo]$. Recall that
$\nablu\eq\partial/{\partial z_u}\pl\leu$. Thus we first study the
effect of differentiating \wrt $z_u$. Employing the identities
  \be  \Frac\partial{\partial z_u}\, \gsO = - \fisu \quad\ {\rm for}\;\
  \so\nE u \qquad{\rm and}
  \qquad \Frac\partial{\partial z_u}\,\hs\zuo = \fius \quad\ {\rm for}\;\
  s\nE u \,, \ee
we derive
  \be  \bearll \Frac\partial{\partial z_u}\, \tommuvo(\vz) \zuo = 
  \Llb -(\muu,H)\, \fisuo + K\,\Frac\partial{\partial z_u}\,\fsO\zuo \Lrb\,
  \tommuvo(\vzo) & {\rm for}\;\ \so\nE u \,, \nxl7
  \Frac\partial{\partial z_u}\, \tommuvo(\vz) \zuo = 
  \Llb \summu (\mus,H)\, \fiuso + K\,\Frac\partial{\partial z_u}\,\fsu\zuo\Lrb\,
  \tommuvo(\vzo) & {\rm for}\;\ \so\eq u \,. \eear \Labl4b
Concerning the commutation with $\leu$, we use
  $    \tommuvs(\vz) L_{-1}^\ssss\eq\ommuvs(L_{-1}^\ssss)\, \tommuvs(\vz) $  
which follows by \Erf TX. Also, according to \Erf Ln we have explicitly
  \be  \ommuvu(\leu) = \leu + \sumlz \summs (\mus{,}H_{-1-\ell})\,
  \Res(t^\ell\fius) + \Frac12\,K\sum_{s,s'=1}^m(\mus{,}\musP)\,\Res(\fius\fiuP)
  \,.  \Labl Lu
The two terms of the \rhs\ can be rewritten with the help of
  \be  \sumlz H^i_{-1-\ell}\,\Res(t^\ell\fius)
      \equiv \sumlz H^i\oT t^{-1-\ell}\,\Res(t^\ell\fius)
  = \sumlz \hi \oT (\fius)_{-1-\ell}\,t^{-1-\ell} = \hi \oT \fius  \ee
and
  \be  \sum_{s,s'=1}^m(\mus{,}\musP)\,\Res(\fius\fiuP)
  = 2\, \summu \frac{(\mus,\muu)}{z_u\mi z_s}  \ee
so as to obtain
  \be  [\tommuvu,\leu] = \llb \ommuvu(\leu)-\leu \lrb\, \tommuvu
  = \Llb \summs (\mus{,}H)\oT\fius + K\,\summu \Frac{(\mus,\muu)}{z_u\mi z_s} 
  \Lrb\, \tommuvu \,.  \ee

Combining this with \Erf4b, we finally get
  \be  [\nablu,\tommuvo(\vz)] =
  -(\muu,H)\ot\fisuo \, \tommuvo(\vzo)+ K\,\Frac\partial{\partial z_u}\,\fsO\zuo
  \, \tommuvo(\vzo) \quad\ {\rm for}\;\ \so\nE u \,,  \Labl8b
while for $\so\eq u$ several terms cancel, leading to
  \be  [\nablu,\tommuvu(\vz)]\zuo =
  - (\muu,H)\ot t^{-1}\, \tommuvu(\vzo) + K\,\Llb \Frac\partial{\partial z_u}\,
  \fsu\zuo - \summu \Frac{(\mus,\muu)}{z_u\mi z_s} \Lrb\, \tommuvu(\vzo)  
  \,.  \Labl8c
This is indeed equivalent to the formula \Erf8a.

\sect{On the implementation of inner automorphisms}\label{AD}

In this appendix we present the calculations which show that the specific
implementation of inner \gsauto s that was described in \erf{primp}
satisfies the relation \Erf3z and explain why this implies that the 
implementation can be consistently chosen at all points of the moduli space 
$\calm_m$.

We consider first the case of a single coroot $\alphav_s$ for some fixed $s$, 
i.e.\ $\mus\eq\alphav_s$ while $\musP\eq0$ for $s'\nE s$. We write
  \be  \Xs \equiv \Xas(\vz) := \eals\oT \fiss + \emals\oT \fiss^{-1}  \labl{Xs}
and
  \be  \Xso := \Xas(\vzo)= \eals\oT \fisso + \emals\oT(\fisso)^{-1} \,.
  \labl{Xso}
We claim that $\Xs$ and $\Xso$ are related by \Erf3y, which in terms of the 
present notations reads
  \be  \exs = \uz\,\exso\,\uzm \,,  \Labl1d
where
  \be  \uz \equiv \uza := \exp\llb -\half\hals\oT\gsos(\vz) \lrb  \Labl uz
with $\gsos$ as defined in \erf{gsos} and $\hals\,{\equiv}\,(\alphav_s,H)$.
To prove \Erf1d, we show that both sides satisfy the same differential
equation; the relation then holds because owing to $\uzo\eq\bfe$ 
both sides are identical at $\vz\eq\vzo$. To obtain the derivative of the
\lhs\ of \erf{1d} we first compute
   \be \bearll  \Frac\partial{\partial z_u}\,\Xs \!\!
   &= (\delta_{u,s}\mi\delta_{u,\so})\, (\eals\ot\fiss^2 + \emals\ot\id) \nxl5
   &= -\half (\delta_{u,s}\mi\delta_{u,\so})\, \ad\Xs(\hals\oT\fiss) 
   \equiv \half (\delta_{u,s}\mi\delta_{u,\so})\, \ad{\hals\ot\fiss}(\Xs) 
   \,,  \eear \ee
where in the transition to the second line we used the formula \Erf12.
It follows immediately that
   \be  \Frac\partial{\partial z_u}\,\Exs           
   = \half (\delta_{u,s}\mi\delta_{u,\so})\, \llb \hals\ot\fiss\,\Exs
   - \Exs\,\hals\ot\fiss \lrb \,.  \Labl2d
As for the \rhs\ of \Erf1d, we have
   \be \bearll  \Frac{\partial\uz}{\partial z_u} \!\!\!
   &= -\half \uz\cdot \hals\oT \Frac\partial{\partial z_u}\,\gsos \nxl5
   &= -\half \uz\cdot \hals\oT \fiss\,(\delta_{u,\so}\mi\delta_{u,s})
   \equiv \half(\delta_{u,s}\mi\delta_{u,\so})\, \hals\oT \fiss\cdot \uz
   \eear \ee
and
   \be  \Frac\partial{\partial z_u}\,\uzm
   = - \uzm\, \Frac{\partial\uz}{\partial z_u}\,\uzm
   = -\half(\delta_{u,s}\mi\delta_{u,\so})\, \uzm\cdot \hals\oT \fiss \,;  \ee
therefore the product $\uz\,\exso\,\uzm$ indeed satisfies the same differential
equation as $\exs$ in \Erf2d.

Next we note that according to the result \Erf B4 and \erf{B15}, we also have
  \be  \expad{(\ipi/2)\Xs}(\hbe\ot f) = \bar w_\alphab(\hbe) \oT f
   - (\alphav,\betab^\Vee)\,K\,\Res\llb \fiss^{-1}\rmdd\fiss\, f \lrb \,.  \ee
Thus in particular
  \be  \bearll \expad{(\ipi/2)\Xs}(\hals\ot f) \!\!\!
  &= - \hals\oT f 
   - (\alphav_s,\alphav_s)\,K\,\Res\llb \fiss^{-1}\rmdd\fiss\, f \lrb\,, \nxl4
  \expad{-(\ipi/2)\Yss}(\hals\ot f) &= - \hals\oT f \,,  \eear \ee
where we introduced
  \be  \Yss := (\eals+\emals) \oT \id \,.  \labl{Yss}
In other words, $\,\eys\,\hals\ot f\eq{-}\hals\ot f\eys$, which implies in 
particular that
  \be  \eys\,\uzm = \uz\,\eys \,,  \Labl3d
while
  \be  \Exs\,\uz = \uzm\,\Exs \cdot \exp\llb{-}\killings\,K\,
  \Res(\fiss^{-1}\rmdd\fiss\,\gsos) \lrb\,,  \labl3j 
which by $\rmdd\fiss\eq{-}\fiss^2$ and
  \be  \Res(\fissp\gsos) = \left\{ \bearll \delta_{\so,s'}\, \ln
  \Frac{z_\so-z_s}{z_\so^\kln-z_s^\kln} & {\rm for}\ s\nE\so\,, \nxl9
  0  & {\rm for}\ s\eq\so   \eear\right. \Labl35
reduces to
  \be  \exs\,\uz = \uzm\,\exs \,.  \labl3k 

Now according to our prescription \erf{primp},
at each individual point $\vz$ in the moduli space
the inner \auto\ $\Exy$ of the \bla\ \gT\ is implemented by
  \be  \ttoalpho(\vz) = \eys\,\exs \,.  \labl{toalpho}
Combining the results \Erf1d, \Erf3d and \Erf3k, we learn that this twisted
intertwiner satisfies
  \be \bearll  \ttoalpho(\vzo) \!\!
  &= \eys\,\exso \nxl3
  &= \eys\,\uzm\,\exs\,\uz  \nxl3
  &= \uzz\,\eys\,\exs = \uzz\,\ttoalpho(\vz) \,.  \eear \Labl3f

We now return to the general case where the components $\muv$ 
are arbitrary elements of the coroot lattice. We note that because of
  \be  \Res\llb \rmdd\gsos\,\gsosp \lrb = 0  \ee
for all values of $s,s'$ any two operators of the form \Erf uz commute 
(when taken at the same point $\vz$ in the moduli space $\calm_m$
of course). Moreover, according
to \erf{omH2}, up to central terms $\uz$ also commutes with the twisted
intertwiner $\ttobeto(\vz)$ for $s'\nE s$ and arbitrary coroots
$\betab^\Vee_{}\,{\equiv}\,\betab^\Vee_{s'}$; more precisely, using the
identity \Erf35 we find
  \be  \bearll \ttobeto(\vz)\,\uz \!\!
  &=\uz\,\ttobeto(\vz)\,\cdot \exp\LLb -\half (\betab^\Vee_{s'},(\alphav_s))\,
    K\,\Res(\fissp\,\gsos) \LRb \nxl6
  &=\uz\,\ttobeto(\vz)\,\cdot \exp\LLb -\half (\betab^\Vee_{s'},(\alphav_s))\,
    \delta_{\so,s'}\,K\, \ln \Frac{z_\so-z_s}{z_\so^\kln-z_s^\kln} \LRb
    \,.  \eear \ee
The final expression will involve the square of the exponential with the
central element, because it is $\uzz$ that we commute to the left. Moreover,
taking into account the chosen ordering of the factors in $\tttommuvo$, such a
term will appear if and only if $s\,{>}\so$. Thus defining $\tttommuvo$ 
according to the prescription \erf{primp}, i.e.\ as
  \be  \tttommuvo(\vz) := \prod_s^\to \Llb \prod_{i_s}^\to
  \eyis\,\exis \Lrb  \labl3i
when $\mus\eq\sum_{i_s}\alphab_{i_s}^\Vee$ for $s\eq\onetome$,
the equation \Erf3f generalizes to
  \be \bearll \tttommuvo(\vzo) = \Uzz\,\tttommuvo(\vz) \cdot \exp\LLb
  -\sumsge (\mub_\so,\mub_s)\,K\, \ln\Frac{z_\so-z_s}{z_\so^\kln-z_s^\kln}
  \LRb  \eear \labl3g
with
  \be  \Uz := \exp\llb -\summse\half(\mus,H)\oT\gsos(\vz) \lrb \,.  \Labl Uz
Using the invariance of the exponent under exchange of $s$ with $\so$, \Erf3g
may also be rewritten as
  \be \bearll \tttommuvo(\vzo) = \Uzz\,\tttommuvo(\vz) \cdot \exp\LLb
  -\half\sumsse (\mub_\so,\mub_s)\,K\, \ln\Frac{z_\so-z_s}{z_\so^\kln-z_s^\kln}
  \LRb \,.  \eear \labl3v

We are now finally in a position to make contact to the relation \Erf3s
that corresponds to the compatibility with the \kzc. First, comparison of our
results with the definition \erf{Inner}\,%
 \futnote{The summation over $s$ in \erf{Inner} now
extends only up to $\me$, owing to our decision to keep one puncture at
$z_m\eq\infty$ when dealing with inner \gsauto s.}
(see also \erf{gsos} and \Erf5u), tells us that we may rewrite \Erf3v 
in the form of \Erf3z, i.e.
  \be  \tttommuvo(\vz) =\Inner\muv(\vz;\vzo)\, \tttommuvo(\vzo) \,.  \labl3l
Moreover, we already know that the implementation is consistent at the
specific point $\vzo$, i.e.\ we have $\tttommuvo(\vzo)\eq\tommuvo(\vzo)$.
Together with the relation \Erf3s between $\tommuvo(\vz)$ and $\tommuvo(\vzo)$
the result \Erf3l therefore implies that in fact
  \be  \tttommuvo(\vz) =\Inner\muv(\vz;\vzo)\, \tommuvo(\vzo) 
  = \tommuvo(\vz)  \ee
at every point $\vz$ in the moduli space $\calm$ and hence that, as claimed,
the implementation is consistent on all of $\calm$.

\vskip3em

\noindent{\bf Acknowledgement}: We would like to thank A.N.\ Schellekens for
helpful discussions and K.\ Gaw\c edzki for pointing out some errors in
an earlier version of the paper.

\newpage \small

 \newcommand\wb{\,\linebreak[0]} \def\wB {$\,$\wb}
 \newcommand\Bi[1]    {\bibitem{#1}}
 \newcommand\Erra[3]  {\,[{\em ibid.}\ {#1} ({#2}) {#3}, {\em Erratum}]}
 \newcommand\BOOK[4]  {{\em #1\/} ({#2}, {#3} {#4})}
 \newcommand\J[5]     { {\sl #5}, {#1} {#2} ({#3}) {#4} }
 \newcommand\inBO[7]  { {\sl #7}, in:\ {\em #1}, {#2}\ ({#3}, {#4} {#5}),
                      p.\ {#6}}
 \def\jf    {J.\ Fuchs}
 \def\adma  {Adv.\wb Math.}
 \def\anif  {Ann.\wb Inst.\wB Fou\-rier}
 \def\anop  {Ann.\wb Phys.}
 \def\aspm  {Adv.\wb Stu\-dies\wB in\wB Pure\wB Math.}
 \def\comp  {Com\-mun.\wb Math.\wb Phys.}
 \newcommand\geap[2] {\inBO{Physics and Geometry} {J.E.\ Andersen, H.\
            Pedersen, and A.\ Swann, eds.} \MD\NY{1997} {{#1}}{{#2}} }
 \def\ijmp  {Int.\wb J.\wb Mod.\wb Phys.\ A}
 \def\joal  {J.\wB Al\-ge\-bra}
 \def\jomp  {J.\wb Math.\wb Phys.}
 \def\lemp  {Lett.\wb Math.\wb Phys.}
 \newcommand\mbop[2] {\inBO{The Mathematical Beauty of Physics}
            {J.M.\ Drouffe and J.-B.\ Zuber, eds.} \WS\Si{1997} {{#1}}{{#2}} }
 \def\nupb  {Nucl.\wb Phys.\ B}
 \newcommand\phgt[2] {\inBO{Physics, Geometry, and Topology}
            {H.C.\ Lee, ed.} \PL\NY{1990} {{#1}}{{#2}} }
 \def\phlb  {Phys.\wb Lett.\ B}
 \def\pnas  {Proc.\wb Natl.\wb Acad.\wb Sci.\wb USA}
 \def\A      {Algebra}
 \def\Ad     {{Amsterdam}}
 \def\alg    {algebra}
 \def\aste   {Ast\'e\-ris\-que}
 \def\Be     {{Berlin}}
 \def\BIR    {{Birk\-h\"au\-ser}}
 \def\Ca     {{Cambridge}}
 \def\class  {classification }
 \def\compac {compactification}
 \def\con    {conformal\ }
 \def\CS     {Chern\hy Si\-mons }
 \def\CUP    {{Cambridge University Press}}
 \def\dimn   {dimension}
 \def\furu   {fusion rule}
 \def\GB     {{Gordon and Breach}}
 \def\ide    {identification}
 \newcommand{\iNBO}[7]{{\sl #7}, in:\ {\em #1} ({#5}), p.\ {#6}}
 \def\Infdim {Infinite-dimensional}
 \def\Intro  {Introduction }
 \def\inv    {invariance}
 \def\kz     {Knizh\-nik\hy Za\-mo\-lod\-chi\-kov }
 \def\kzbe   {Knizh\-nik\hy Za\-mo\-lod\-chi\-kov\hy Ber\-nard equation}
 \def\MD     {{Marcel Dekker}}
 \def\modinv {modular invarian}
 \def\modinvt{modular invariant}
 \def\NH     {{North Holland Publishing Company}}
 \def\nn     {$N\,{=}\,2$ }
 \def\NY     {{New York}}
 \def\parfu  {partition function}
 \def\PL     {{Plenum}}
 \def\Q      {Quantum\ }
 \def\qg     {quantum group}
 \def\qzn    {quantization}
 \def\Rep    {Representation}
 \def\Si     {{Singapore}}
 \def\suco   {superconformal }
 \def\SV     {{Sprin\-ger Verlag}}
 \def\sym    {symmetry}
 \def\syms   {sym\-me\-tries}
 \def\WZ     {Wess\hy Zu\-mino }
 \def\wzw    {WZW\ }
 \def\va     {Virasoro algebra}
 \def\WI     {{Wiley Interscience}}
 \def\WS     {{World Scientific}}
 \def\ybe    {Yang\hy Bax\-ter equation}

 \end{document}